\documentclass[twocolumn,showpacs,preprintnumbers,amsmath,amssymb,aps,prl]{revtex4-1}

\usepackage{graphicx}
\usepackage{dcolumn}
\usepackage{bm}
\usepackage{amsmath,amssymb}
\usepackage{epstopdf}
\usepackage{psfrag}
\usepackage[T1]{fontenc}
\usepackage{hyperref}
\usepackage{tabularx}
\usepackage{color}
\usepackage{float}
\newcommand\numberthis{\addtocounter{equation}{1}\tag{\theequation}}

\newcolumntype{Y}{>{\centering\arraybackslash}X}
\begin{document}

\preprint{}

\title{Structural and thermal transport properties of ferroelectric domain walls in GeTe from first principles}
\author{{\DJ}or{\dj}e Dangi{\'c}\textsuperscript{1,2}}
\email{djordje.dangic@tyndall.ie}
\author{\'Eamonn D. Murray\textsuperscript{3}}
\author{Stephen Fahy\textsuperscript{1,2}}
\author{Ivana Savi\'c\textsuperscript{2}}
\email{ivana.savic@tyndall.ie}
\affiliation{\textsuperscript{\normalfont{1}}Department of Physics, University College Cork, College Road, Cork, Ireland}
\affiliation{\textsuperscript{\normalfont{2}}Tyndall National Institute, Dyke Parade, Cork, Ireland}
\affiliation{\textsuperscript{\normalfont{3}}Department of Physics and Department of Materials, Imperial College London, London SW7 2AZ, UK}

\date{\today}

\begin{abstract}

Ferroelectric domain walls are boundaries between regions with different polarization orientations in a ferroelectric material. Using first principles calculations, we characterize all different types of domain walls forming on ($11\bar{1}$), ($111$) and ($1\bar{1}0$) crystallographic planes in thermoelectric GeTe. We find large structural distortions in the vicinity of most of these domain walls, which are driven by polarization variations. We show that such strong strain-order parameter coupling will considerably reduce the lattice thermal conductivity of GeTe samples containing domain walls with respect to single crystal. Our results thus suggest that domain engineering is a promising path for enhancing the thermoelectric figure of merit of GeTe.  

\end{abstract}


\maketitle

\section{I. Introduction} 

Thermoelectric materials can convert heat into electrical power or, in reverse, cool devices using electrical current. GeTe is one of the most efficient thermoelectric materials currently known \cite{GeTe2,gete-jacs,yaniv-gete-jap-16, biswas-gete-rev, GeTe3, GeMnTe, GeTe1, GeTePNAS, Li2018, Zhang2018, Dong2019, Wu2019, Hong2019}. 
One reason for this is the low lattice thermal conductivity \citep{Ivana1, CampiGeTe}, which is the result of its proximity to the ferroelectric phase transition mediated by soft transverse optical (TO) phonon modes. These soft phonon modes strongly couple with heat carrying acoustic phonons, disrupting their flow and leading to the low lattice thermal conductivity \cite{Ivana1}. Secondly, certain electronic band structure properties of GeTe (such as high valence band degeneracy, valence band convergence, relatively small band mass \cite{GeTe3}) further improve its thermoelectric performance.

Below the Curie temperature, GeTe samples may contain regions with different polarization orientations, which are known as ferroelectric domains \cite{DWGeTe1, DWGeTe2, RashbaGeTe, ActaMatGeTeDW, SNYKERS1972,PolkingGeTeDW,Kim2019, KrignerGeTeDW}. Ferroelectric domains in a material are separated by regions of varying polarization called domain walls (DWs) \cite{RevModPhysNanoDom,MainBiFeO3,InsulatingYMnO3,BiFeO3PRL,CondTTAPL,Fermilvlperovskites}. Influence of DWs on the thermoelectric transport properties of GeTe and ferroelectric materials in general has not been much investigated. Here we investigate whether DWs, like other types of interfaces, could suppress the lattice thermal conductivity of GeTe, as reported in ferroelectric oxides~\cite{Mante1971,Weilert1993,Mielcarek2001,Hopkins2013,Li2014,Ihlefeld2015,DWThermCond1,DWThermCond2,Foley2018}.

\begin{figure}[h]
\begin{center}
  \includegraphics[width = 0.9\linewidth]{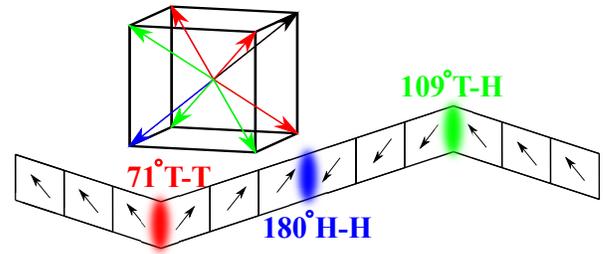}
\caption{An example of the domain structure in a ferroelectric material. Different polarization directions in a single domain are color coded (top) and the boundaries between these domains correspond to different types of domain walls (bottom). Domain wall boundary for this configuration of domains is one of the faces of the cube, ($001$) plane. Tail-to-tail, head-to-head and tail-to-head domain walls are labeled as T-T, H-H and T-H, respectively.}
\label{fig1}
\end{center}
\end{figure}

Below $\sim 600\--700$~K, GeTe crystallizes in a rhombohedral structure, characterized by the Te internal atomic displacement along the [111] direction, which represents the order parameter. GeTe crystal can be described using the primitive rhombohedral unit cell with two atoms, the hexagonal unit cell with six atoms, the pseudocubic unit cell with eight atoms etc. The sides of each of these unit cells can be the host planes for DWs, forming different types of DWs in relation to the DW plane e.g. (11$\bar{1}$), (1$\bar{1}$0), (111) and (001) DWs (these are Miller indices of the planes in the pseudocubic unit cell). Each of these planes can host different DWs depending on the angle between polarization vectors in neighboring domains: 39$^{\circ}$, 141$^{\circ}$ and 180$^{\circ}$ (11$\bar{1}$) DWs, 180$^{\circ}$ (1$\bar{1}$0) and (111) DWs, and 71$^{\circ}$, 109$^{\circ}$ and 180$^{\circ}$ (001) DWs. Depending on the orientation of polarization vectors with respect to the DW plane, there are two additional types of 39$^{\circ}$ and 180$^{\circ}$ (11$\bar{1}$) DWs, 180$^{\circ}$ (111) DWs, and 71$^{\circ}$ and 180$^{\circ}$ (001) DWs: head-to-head (H-H) and tail-to-tail (T-T). 141$^{\circ}$ (11$\bar{1}$), 180$^{\circ}$ (1$\bar{1}$0) and 109$^{\circ}$ (001) DWs have a head-to-tail (H-T) (or equivalently, tail-to-head, T-H) polarization orientation. Some of (001) DWs are illustrated in Fig.~\ref{fig1}. All described H-H and T-T DWs are charged i.e.~they have induced bound charge due to polarization discontinuity at the DW, while all considered H-T DWs are neutral (the polarization component perpendicular to the DW plane does not change, hence there is no bound charge). Considering the nature of polarization change across DWs, they can have Ising character, where polarization only changes in magnitude, and Bloch or N\'eel character if polarization rotation occurs in the plane parallel or perpendicular to the DW plane, respectively \cite{Vanderbilt1}.

While available experimental data agree that the herringbone domain structures are present in GeTe samples, they disagree on the predominant types of domain walls. Ref.~\cite{ActaMatGeTeDW} reported (001), (1$\bar{1}$0) and (11$\bar{1}$) DWs, while other studies \cite{DWGeTe1, PolkingGeTeDW} found (001) and (110) DWs. The most recent work  \cite{Kim2019} claims that the herringbone structure is bounded by the (110) and (111) planes, stabilizing (11$\bar{1}$) DWs after doping GeTe with Sb and Si. In the light of these contradictory experimental findings, computational investigation of ferroelectric DWs in GeTe gains in importance.

In this paper, we characterize all the described types of (11$\bar{1}$), (111) and (1$\bar{1}$0) domain walls in GeTe from first principles. We calculate structural properties, such as DW energy and width, polarization profile, and local structure distortions, of each of those DWs. We find that all of the investigated DWs have Ising - N{\'e}el character, except ($111$) DWs that are purely Ising. Strong strain-order parameter coupling is present at most of these DWs, which amplifies strong acoustic-soft TO mode coupling that exists in domains. As a result, the lattice thermal conductivity of GeTe samples incorporating ferroelectric domains can be significantly lower than that of single crystal in the direction perpendicular to the DW plane. These findings demonstrate the potential of domain wall engineering for improved thermoelectric performance.    

\section{Construction and relaxation of domain walls} 

The low temperature value of the rhombohedral angle in GeTe is 57.825$^{\circ}$ \cite{GeTeBo}, which is well captured with our density functional theory (DFT) calculation yielding 57.776$^{\circ}$. Such a large value of the rhombohedral distortion makes the task of constructing GeTe supercells that contain DWs more difficult than that for cubic materials  \cite{Vanderbilt1, DFTBiFeO31, Dieguez2013, DFTBiFeO32, GONG20189}, since we need to realistically represent twinning due to lattice orientation mismatch at the DW boundary. Twinning does not occur for 180$^{\circ}$ DWs and their construction is straightforward.

\begin{figure}[h]
\begin{center}
\includegraphics[width = 0.8\linewidth]{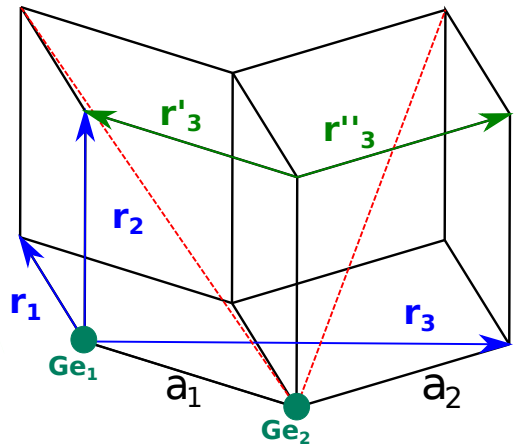}
\caption{Geometry of GeTe domain structure containing 39$^{\circ}$ or 141$^{\circ}$ ($11\bar{1}$) domain walls for the case where domains are one unit cell long. Blue lines are the unit cell vectors of the domain structure, while green vectors represent the third primitive lattice vectors of individual domains. Red lines represent polarization directions in different domains. The positions of Ge atoms are labeled as Ge$_1$ and Ge$_2$. The lattice constants of the two primitive unit cells that constitute this supercell are labeled as $a_{1}$ and $a_{2}$.}
\label{fig2}
\end{center}
\end{figure}

We construct supercells containing 39$^{\circ}$ and 141$^{\circ}$ ($11\bar{1}$) twinned DWs as follows. The primitive unit cell of rhombohedral GeTe is defined by the translation vectors:
\begin{align*}
\vec{r}_{1} &= a(b,0,c), \\
\vec{r}_{2} &= a(-\frac{b}{2},\frac{b\sqrt{3}}{2},c), \numberthis \\
\vec{r'}_{3} &= a(-\frac{b}{2},-\frac{b\sqrt{3}}{2},c),
\label{eq1}
\end{align*}
where $a$ is the lattice constant, $b = \sqrt{2(1-\cos\theta)/3}$, $c = \sqrt{(1+2\cos\theta)/3}$, and $\theta$ is the angle between the primitive lattice vectors. The atomic positions in this structure are taken to be: Ge (0.0,0.0,0.0) and Te (0.5 + $\tau$, 0.5 + $\tau$, 0.5 + $\tau$) in reduced coordinates. We choose one of the crystallographic planes in the primitive unit cell to be our DW boundary, for example ($11\bar{1}$). In this case we keep the first and second lattice vectors unchanged prior to structural relaxation and they are identical in both domains.  We calculate the third lattice vector of the second domain $\vec{r''}_{3}$ using the fact that ($11\bar{1}$) plane is the mirror plane of our domain structure, see Fig.~\ref{fig2}. The third lattice vector $\vec{r}_{3}$ for the entire structure is defined as $\vec{r}_3=N(\vec{r''}_3-\vec{r'}_3)/2$, where $N$ is the number of primitive unit cells in the supercell, which contains two DWs: 39$^{\circ}$ H-H and T-T DWs, or 141$^{\circ}$ H-T and T-H DWs. 

After constructing the supercells described above, we relax the atomic positions and structure using DFT. First we relax the positions of Te atoms, keeping Ge atoms and the global structure (the unit cell vectors $\vec{r}_{1}$, $\vec{r}_{2}$ and $\vec{r}_{3}$) fixed. In this case, forces after relaxation are around 10$^{-4}$ eV/\AA {} inside the domains and can be as large as 0.1 eV/\AA {} at the DW. The second step is the relaxation of the local structure through the relaxation of Ge atomic positions and the supercell lattice vectors, along with further optimization of Te atomic positions. After this step, atomic forces are lower than 10$^{-6}$ eV/\AA {} even for atoms at the DW. We used these structures for the calculation of DW energies and widths, and local structure distortions.

Next we define local structure parameters, which are descriptive of one primitive unit cell within the constructed supercell. We describe the local lattice constant for the $i$th primitive cell away from the DW as the distance between two neighboring Ge atoms:
\begin{align*}
a_{i} = |\vec{a}_{i}| = |\vec{r}_{Ge,i} - \vec{r}_{Ge,i+1}|. \numberthis
\end{align*}
The local rhombohedral angle for the same primitive cell is calculated from the scalar product of $\vec{a}_{i}$ with the first supercell translation vector $\vec{r}_{1}$ (using $\vec{r}_{2}$ yields the same result):
\begin{align*}
\theta _{i} = \arccos{\frac{\vec{a}_{i}\cdot\vec{r}_{1}}{|\vec{a}_{i}||\vec{r}_{1}|}}. \numberthis
\label{eq3}
\end{align*}

\begin{figure}[h]
\begin{center}
\includegraphics[width = 0.8\linewidth]{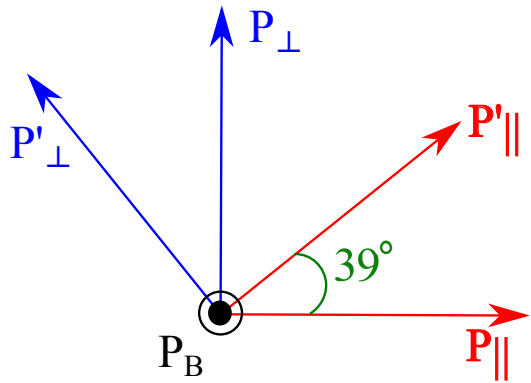}
\caption{Polarization directions inside each domain for GeTe structures containing 39$^{\circ}$ or 141$^{\circ}$ ($11\bar{1}$) domain walls. Polarization directions in two neighboring domains are labeled as primed and non-primed.  $P_{||}$ is the direction along the trigonal axis. $P_{B}$  is the direction normal to the plane of the trigonal axes in neighboring domains and corresponds to the Bloch character of polarization. The third direction, $P_{\perp}$, is perpendicular to the other two directions and quantifies the N{\'e}el character of polarization.}
\label{fig3}
\end{center}
\end{figure}

We define the local polarization vector for each primitive cell as the vector between Te atom and the high symmetry point (0.5,0.5,0.5) inside the same primitive cell and normalize its value so that the polarization magnitude along the trigonal axis inside the domain is one. Polarization profiles are taken along different directions illustrated in Fig. \ref{fig3}. The first direction is along the trigonal axis inside a particular domain, $P_{\parallel}$ (red color in Fig. \ref{fig3}). This direction changes from one domain to another (from $P_{\parallel}$ to $P'_{\parallel}$). The second direction is along the vector normal to the plane defined by the trigonal axes in neighbouring domains $P_B$ (black vector in Fig. \ref{fig3}), which corresponds to the Bloch component of polarization. The third direction is chosen to form the orthogonal coordinate system with the first two directions inside individual domains (blue vectors labeled as $P_{\perp}$ and $P'_{\perp}$  in Fig. \ref{fig3}), representing the N{\'e}el components of polarization.

To extract domain wall widths, we fit the polarization profiles along the trigonal axes to the expression:
\begin{align*}
p(d) = P_{0}\tanh{\frac{2(d-d_{0})}{w}}. \numberthis
\label{eq4}
\end{align*}
Here $w$ is the DW width, $d_{0}$ is the position of the DW boundary and $P_{0}$ is the polarization value inside domains.

Domain wall energies are calculated as:
\begin{align*}
E_{DW} = \frac{E_{1} - E_{0}}{2S}, \numberthis
\label{eq2} 
\end{align*}
where $E_{1}$ is the total energy of the relaxed domain structure, $E_{0}$ is the total energy of bulk GeTe with the same number of atoms as the supercell containing DWs, and $S$ is the area of the DW boundary. Due to periodic boundary conditions, our supercells with a H-H DW must also contain a T-T DW. Therefore we can only calculate an average domain wall energy for a certain DW angle.
 
In the case of 180$^{\circ}$ ($11\bar{1}$) DWs, there is no twinning at the domain boundary and the construction of supercells containing these DWs is trivial. The definition of the polarization directions for these supercells is somewhat ambiguous, since the polarization vectors in neighboring domains are collinear. We choose the supercell in which the polarization directions inside domains are along the $z$ Cartesian axis. We define the Bloch component of polarization along the direction perpendicular to the $z$ axis and the vector of the DW boundary. This allows us to define the N{\'e}el component along the direction perpendicular to the $z$ axis and the Bloch component.

We construct supercells incorporating 180$^{\circ}$ ($111$) and ($1\bar{1}0$) DWs from the hexagonal unit cell of GeTe. The hexagonal unit cell is defined with the following set of lattice vectors:
\begin{align*}
\vec{h}_{1} &= a(\frac{\sqrt{3}b}{2},-\frac{3b}{2},0), \\
\vec{h}_{2} &= a(\frac{\sqrt{3}b}{2},\frac{3b}{2},0), \numberthis \\
\vec{h}_{3} &= a(0,0,3c).
\end{align*}
The definition of parameters $a$, $b$ and $c$ are the same as in the case of the rhombohedral cell. The positions of atoms in this unit cell are: Ge ((0.0,0.0,0.0), (2/3,1/3,1/3), (1/3,2/3,2/3)) and Te ((0.0,0.0,0.5+$\tau$), (2/3,1/3,5/6+$\tau$), (1/3,2/3,1/6+$\tau$)). ($111$) DWs are perpendicular to the trigonal axis, while ($1\bar{1}0$) DW boundary contains the trigonal axis. For ($111$) DWs, the trigonal axis is oriented along the $z$ Cartesian axis, and polarization directions correspond to the Cartesian axes.
For ($1\bar{1}$0) DWs, the N{\'e}el component of polarization is the vector of the DW plane, while the Bloch component is perpendicular to it and the trigonal axis.

(001) DWs are constructed in a similar manner as ($11\bar{1}$) DWs, but using the pseudocubic unit cell (conventional rocksalt structure) vectors:
\begin{align*}
\vec{p}_{c1} &= a(-2b, 0, c), \\
\vec{p}_{c2} &= a(b, -b\sqrt{3}, c), \numberthis \\
\vec{p}_{c3} &= a(b, b\sqrt{3}, c).
\end{align*}
Parameters $a$, $b$ and $c$ are the same as for the rhombohedral cell. We have tried performing relaxation of these domain walls as well. The relaxation of these structures proved to be very computationally expensive, mostly because these domain walls have approximately four times more atoms per domain length compared to ($11\bar{1}$) DWs. In the case of charged DWs, polarization discontinuity induced bound charge is larger at the (001) DWs making them harder to relax. However, we expect similar structural and electronic properties for (001) DWs as for ($11\bar{1}$) and (111) DWs.
  
\section{Technical details}

DFT calculations were performed using the plane wave basis set, the generalized gradient approximation with Perdew-Burke-Ernzerhof parametrization  (\textsc{GGA-PBE}) for the exchange-correlation potential~\cite{GGAPBE} and Hartwigsen-Goedecker-Hutter (HGH) pseudopotentials \cite{HGHpseudo} as implemented in the  \textsc{ABINIT} code \cite{ABINIT,ABINIT2}.  We used the energy cutoff of 16 Ha for plane waves in all cases. We performed a convergence study of the DW widths and energies with respect to the domain size for all considered DWs (see Supplementary Material). We carried out DFT calculations on $1\times 1\times N$ supercells containing ($11\bar{1}$) DWs, where $N$ is 32 for 39$^{\circ}$ and 141$^{\circ}$ DWs (64 atoms) and 40 for 180$^{\circ}$  DWs (80 atoms).  We used a $4 \times 4 \times 1$ \textbf{k}-point grid for the Brillouin zone sampling of the electronic states of ($11\bar{1}$) DWs. For ($111$) and ($1\bar{1}0$) DWs, we used $1\times 1\times 24$ and $24\times 1\times 1$ supercells formed from the hexagonal unit cell (144 atoms). We used $4\times 4\times 1$ and $1\times 12 \times 4$ \textbf{k}-point grids for sampling the Brillouin zone for (111) and ($1\bar{1}0$) DWs, respectively. We used "cold smearing" for electronic states \cite{MarzariPhD} due to the existence of metallic states in some of the structures. All calculations were done excluding spin-orbit coupling.

\section{($\mathbf{11\bar{1}}$) domain walls}

The domain wall energies and widths of $(11\bar{1})$ DWs are presented in Table \ref{tb1}. \footnote{For 39$^{\circ}$ tail-to-tail domain wall, polarization values along the trigonal axis in one of the domains were taken as negative to obtain the $\tanh(x)$ dependence of polarization, see Fig.~\ref{fig2}(a).} The energy cost of DW formation is the largest for 39$^{\circ}$ DWs, and the lowest for 180$^{\circ}$ DWs. Compared to BaTiO$_{3}$ neutral DWs~\cite{Vanderbilt1}, GeTe DWs can have up to 100 times larger DW energies. However, compared to charged DWs in perovskite materials \cite{Vanderbiltpbti03, DFTBiFeO31}, DWs in our calculations have comparable energies. GeTe $(11\bar{1})$ DW energies and widths exhibit a few obvious trends. Charged DWs (39$^{\circ}$ and 180$^{\circ}$) usually have larger energies with respect to the neutral one (141$^{\circ}$). Twinning also gives a large contribution to the DW energy (compare the DW energies of twinned 39$^{\circ}$ and 141$^{\circ}$ DWs with those of 180$^{\circ}$ DWs).

\begin{table}[h]
\begin{center}
\begin{tabularx}{0.5\textwidth}{ c | c | c | Y  }
\hline \hline
 &  H-H width [\AA] & T-T width [\AA] &  Average DW energy [mJ/m$^2$]  \\ \hline
 39$^{\circ}$ DW  & 3.4   & 4.4 & 547 \\ \hline
 180$^{\circ}$ DW & 13.4  & 14.8  & 376  \\ \hline
 & H-T width [\AA]& T-H width [\AA] &  Average DW energy [mJ/m$^2$] \\ \hline 
 141$^{\circ}$ DW & 4.0 & 4.0 & 404  \\ \hline
 \hline 
\end{tabularx}
\end{center}
\caption{Domain wall (DW) widths and energies for $(11\bar{1})$ DWs. H-H and T-T denote head-to-head and tail-to-tail DWs, respectively. H-T and T-H denote head-to-tail and tail-to-head DWs, respectively.}
\label{tb1}
\end{table}

\begin{figure}[h!]
\begin{center}
\includegraphics[width = 0.9\linewidth]{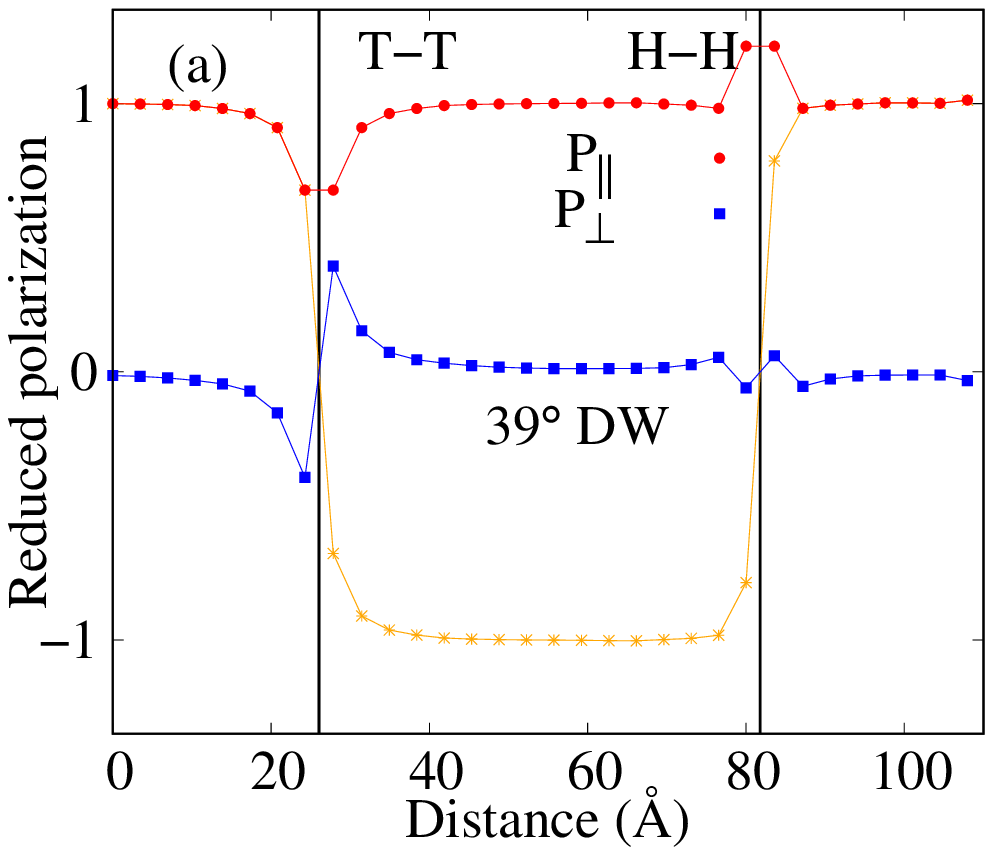}
\includegraphics[width = 0.9\linewidth]{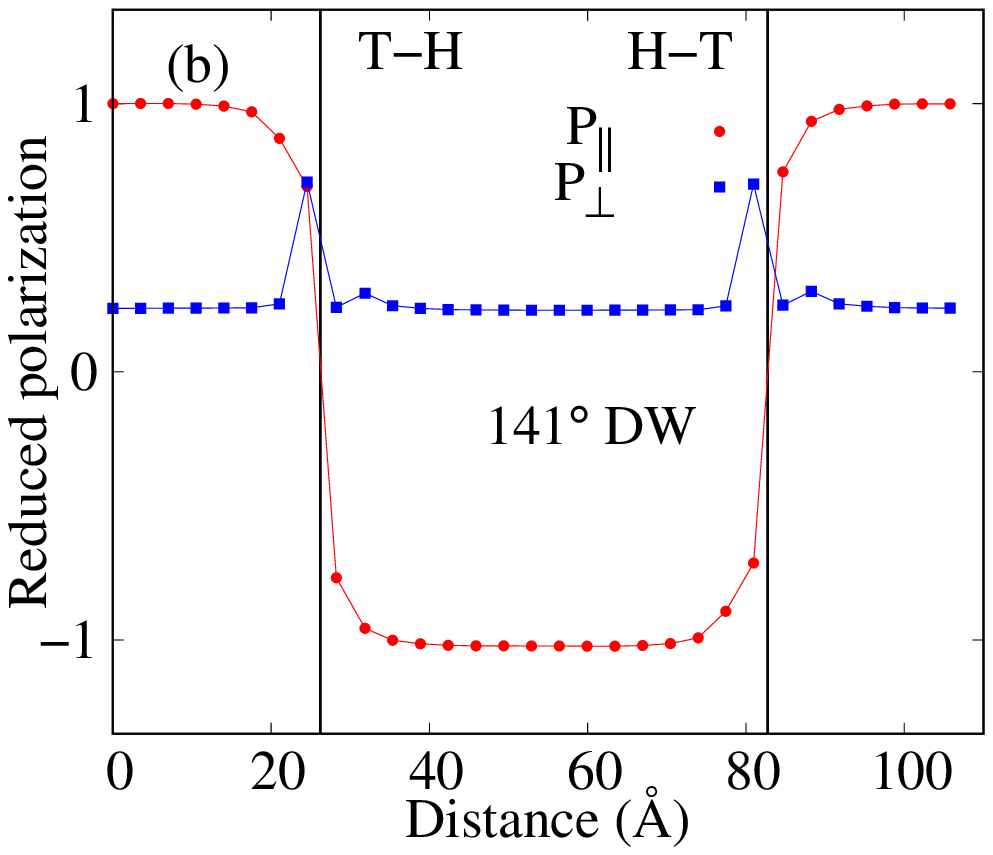}
\includegraphics[width = 0.9\linewidth]{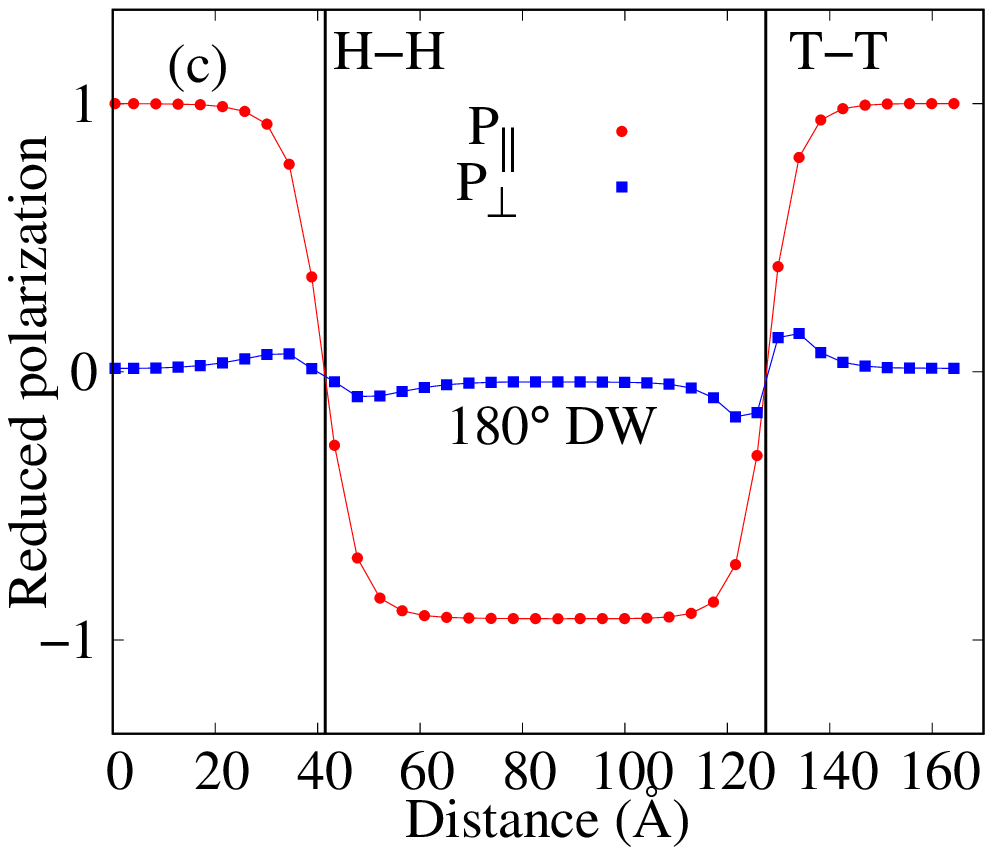}
\caption{Polarization profiles for GeTe structures containing (a) 39$^{\circ}$, (b) 141$^{\circ}$ and (c) 180$^{\circ}$ ($11\bar{1}$) domain walls (DWs). Red line represents polarization in the direction of trigonal axes of each domain (P$_{||}$), and blue line red shows the N{\'e}el component of polarization (P$_{\perp}$). DW boundaries are indicated by black vertical lines and labeled T-T for tail-to-tail, H-H for head-to-head, H-T for head-to-tail and T-H for tail-to-head DWs. For 39$^{\circ}$ T-T DW, P$_{||}$ values in one of the domains are plotted as negative (orange line) to aid visualization.
}
\label{fig4}
\end{center}
\end{figure}

We now discuss the polarization profiles along GeTe structures containing (11$\bar{1}$) domain walls. Bulk GeTe has the spontaneous polarization of 63 $\mu \text{C/cm}^{2}$, which is similar to that of perovskite materials \cite{Vanderbiltpbti03, DFTBiFeO31, Vanderbilt1}. The polarizations along the trigonal axis ($P_{||}$) and the N{\'e}el component of polarization ($P_{\perp}$) for (11$\bar{1}$) DWs are given in Fig.~\ref{fig4}. For all these DWs, the Bloch component of polarization is zero. The N{\'e}el character is stronger for T-T DWs with respect to H-H DWs with the same polarization angle. 141$^{\circ}$ DW has the strongest N{\'e}el character and 180$^{\circ}$ DWs have the strongest Ising character. Both 141$^{\circ}$ DWs have the same polarization profiles since they are of the H-T type, in contrast to 39$^{\circ}$ and 180$^{\circ}$ DWs.

\begin{figure}[h!]
\begin{center}
\includegraphics[width = 0.9\linewidth]{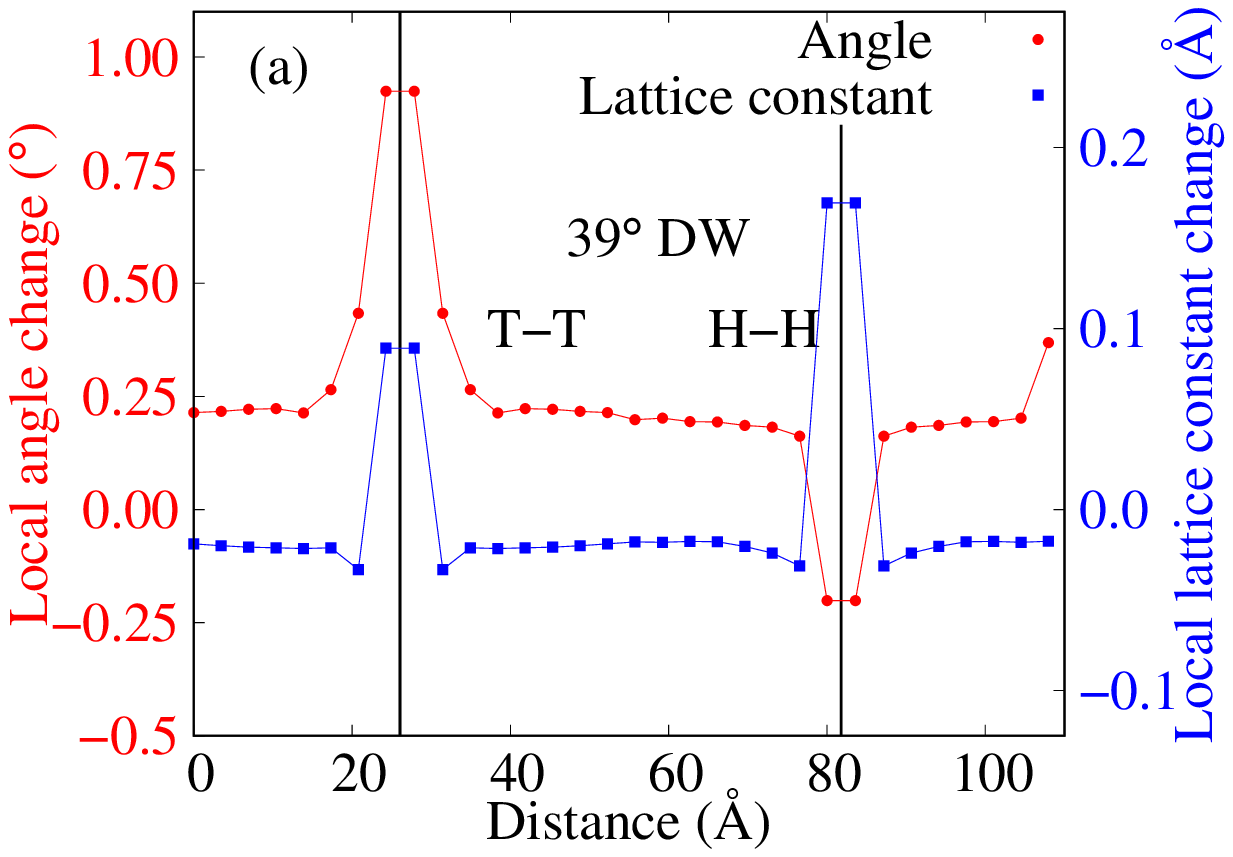}
\includegraphics[width = 0.9\linewidth]{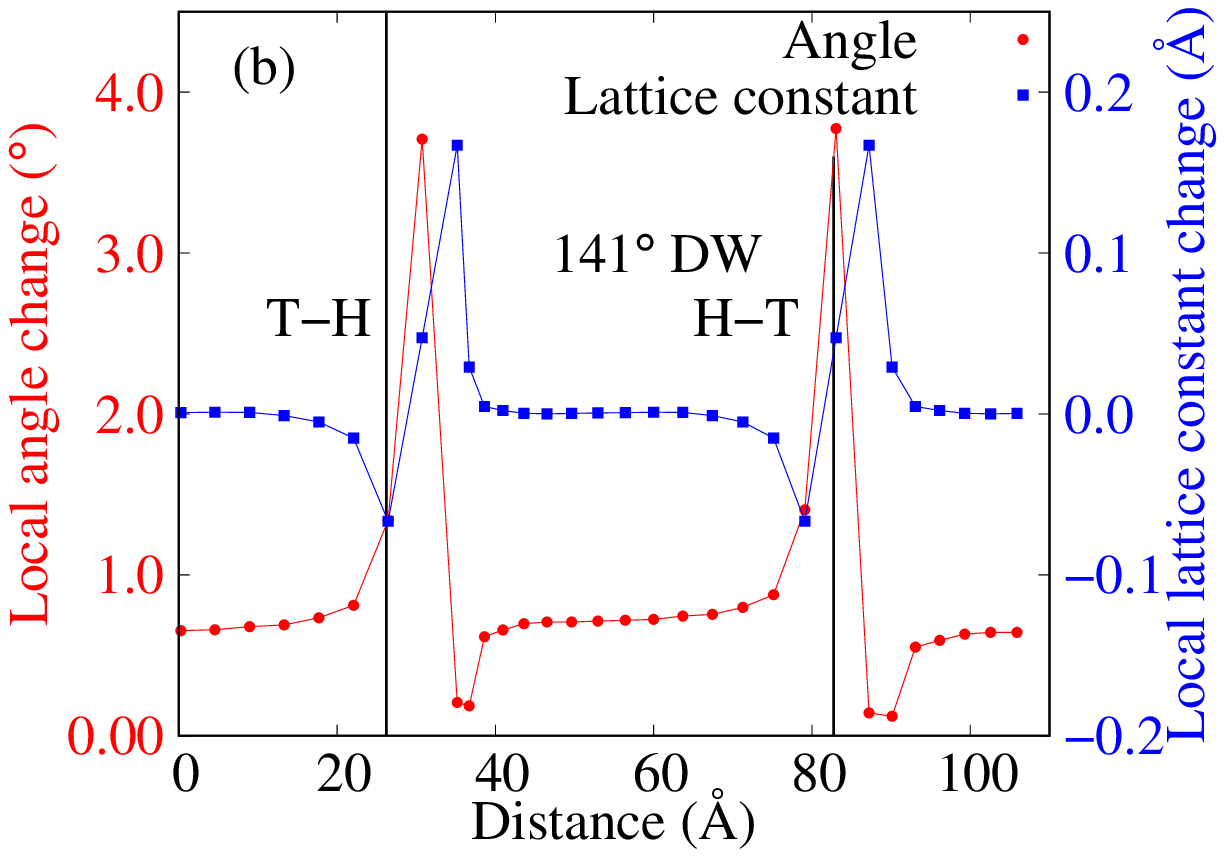}
\includegraphics[width = 0.9\linewidth]{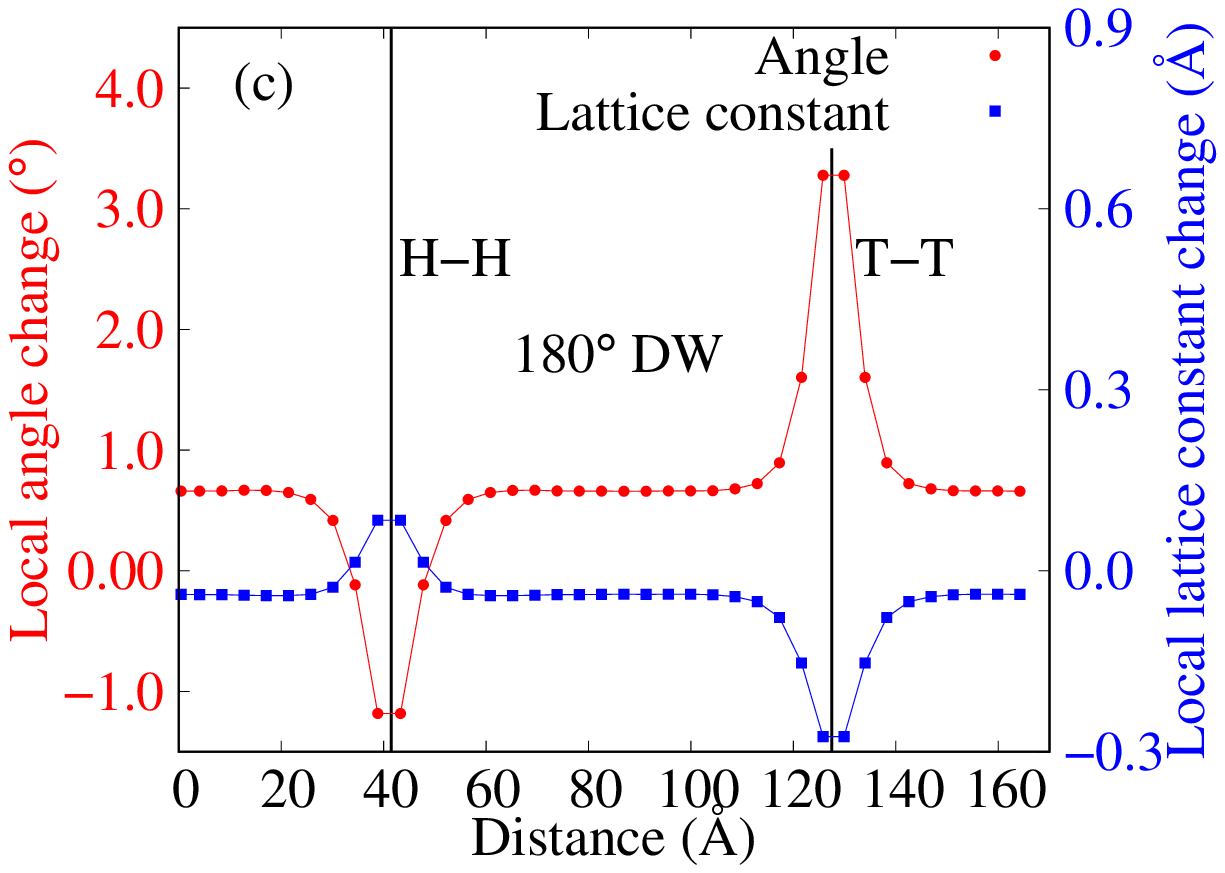}
\caption{Local lattice constant (blue line) and local rhombohedral angle (red line) for GeTe structures containing (a) 39$^{\circ}$, (b) 141$^{\circ}$ and (c) 180$^{\circ}$ ($11\bar{1}$) domain walls (DWs). DW boundaries are indicated by black vertical lines and labeled T-T for tail-to-tail, H-H for head-to-head, H-T for head-to-tail and T-H for tail-to-head DWs.} 
\label{fig5}
\end{center}
\end{figure}

Next we illustrate large local structural distortions in the vicinity of (11$\bar{1}$) DWs. They arise due to the fact that the rhombohedral angle of GeTe is considerably different from the cubic value of 60$^{\circ}$. The local lattice constant and rhombohedral angle changes for supercells incorporating (11$\bar{1}$) DWs with respect to the bulk GeTe values are shown in Fig.~\ref{fig5}. The structural distortions are the smallest in the case of 39$^{\circ}$ DWs, and are considerably larger for 141$^{\circ}$ and 180$^{\circ}$ DWs. We note that, even in the middle of each domain, there is a slight renormalization of the lattice constant and rhombohedral angle compared to the bulk values~\cite{Mi}. This is probably due to the small size of domains that do not perfectly screen the depolarizing field. Also, we point out the asymmetry of structural distortions for 141$^{\circ}$ DW with respect to the DW boundary, which is non-existent in other four types of ($11\bar{1}$) DWs, due to the difference in geometry. 

It is interesting to compare the trends related to local structural changes near DWs to those observed in single crystalline GeTe near the ferroelectric phase transition. In GeTe undoped single crystal, the lattice constant and the internal atomic displacement decrease as the material approaches the phase transition with increasing temperature, while the angle increases~\cite{main, newmain, Mi}. For 39$^{\circ}$ H-H DW, the local lattice constant and polarization along the trigonal axis increase while the local rhombohedral angle decreases closer to the DW. Consequently, the local structure of 39$^{\circ}$ H-H DW exhibits the opposite trends to that of the single crystal near the phase transition. 180$^{\circ}$ T-T DW displays the same trends for the local structure as the single crystal near the phase transition, with decreasing lattice constant and polarization along the trigonal axis and increasing angle closer to the DW.

\section{(111) and (1$\mathbf{\bar{1}}$0) domain walls}

Table~\ref{tb2} shows the domain wall energies and widths of 180$^{\circ}$ (111) and $(1\bar{1}0)$ DWs. Their DW widths are larger compared to $(11\bar{1})$ DWs. We note that there is a difference in the DW widths for individual H-T and T-H (1$\bar{1}$0) DWs. It is unclear whether this is due to the finite size of domains, or there is a symmetry breaking we are not aware of. (1$\bar{1}$0) DW has the smallest energy among all investigated DWs. This is because (1$\bar{1}$0) DW is neutral and its electrostatic energy is small, its N{\'e}el component of polarization is small and there is no twinning at the DW boundary. On the other hand, (111) DWs have the highest energy among all considered DWs and this is mostly due to a large depolarization field caused by bound charge at the DW boundaries.

\begin{table}[h]
\begin{center}
\begin{tabularx}{0.5\textwidth}{ c | Y | Y | Y | Y }
\hline \hline
 & H-H width [\AA] & T-T width [\AA] & \multicolumn{2}{Y}{Average DW energy [mJ/m$^2$]} \\ \hline
 (111) DW & 19.4  & 22.6  & \multicolumn{2}{c}{686}   \\ \hline
 & H-T width [\AA]& T-H width [\AA] &  \multicolumn{2}{Y}{Average DW energy [mJ/m$^2$]} \\ \hline 
 (1$\bar{1}$0) DW & 9.7 & 8.4 & \multicolumn{2}{c}{25} \\ \hline
 \hline 
\end{tabularx}
\end{center}
\caption{Domain wall (DW) widths and energies for 180$^{\circ}$ (111) and $(1\bar{1}0)$ DWs. H-H and T-T denote head-to-head and tail-to-tail DWs, respectively. H-T and T-H denote head-to-tail and tail-to-head DWs, respectively. }
\label{tb2}
\end{table} 

Although our results suggest that ($1\bar{1}0$) DW is much more energetically favorable than other DWs considered, we stress that this study is carried out on perfect crystals, without any imperfections. Including vacancies or interstitial atoms could considerably change the energetics of particulars domains, making other types of DWs more stable. This is somewhat confirmed by a recent experimental study \cite{Kim2019}, which shows that including impurities stabilizes (11$\bar{1}$) DWs.

\begin{figure}[h!]
\begin{center}
\includegraphics[width = 0.9\linewidth]{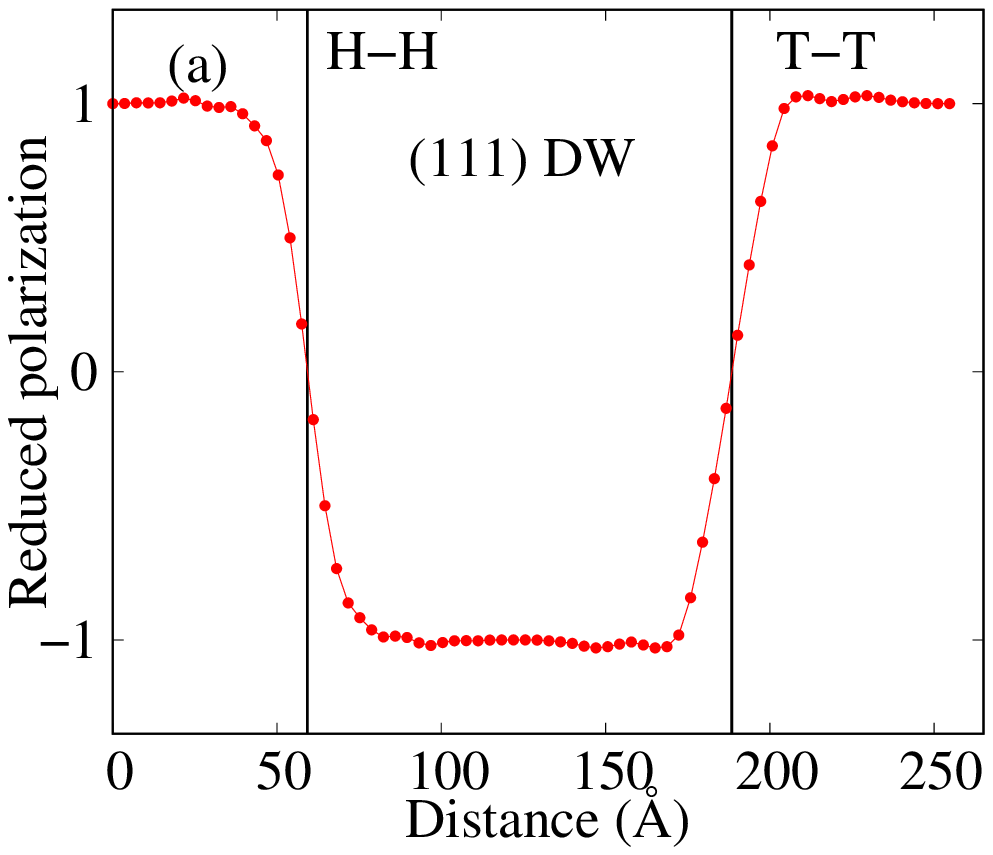}
\includegraphics[width = 0.9\linewidth]{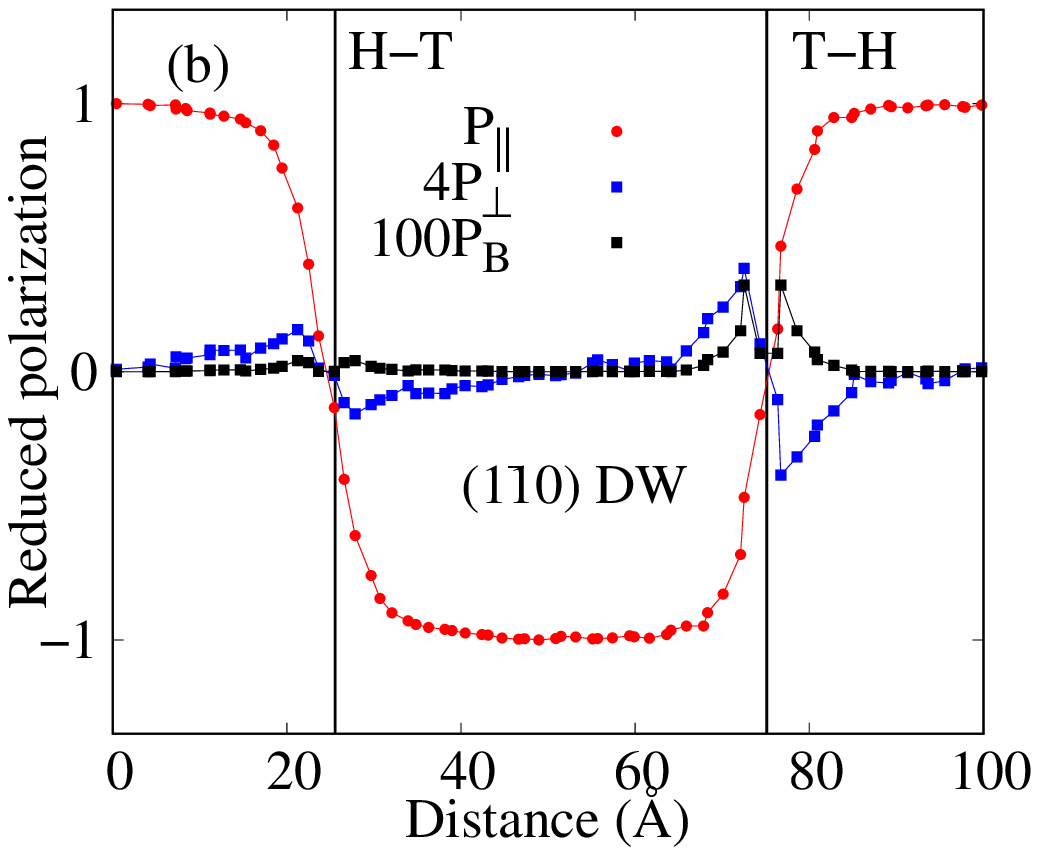}
\caption{Polarization profiles for GeTe structures containing (a) (111) and (b) (1$\bar{1}$0) domain walls (DWs). Red line represents polarization in the direction of the trigonal axis (P$_{||}$), while blue and black lines represent the N{\'e}el (P$_{\perp}$) and Bloch ($P_B$) components of polarization (multiplied by constant values to make them visible on the graph). DW boundaries are indicated by black vertical lines and labeled T-T for tail-to-tail, H-H for head-to-head, H-T for head-to-tail and T-H for tail-to-head DWs.}
\label{fig6}
\end{center}
\end{figure}

The polarization profiles for (111) and ($1\bar{1}0$) DWs are given in Fig. \ref{fig6}. (111) DWs walls have pure Ising character, exhibiting only changes of the magnitude of polarization and not of the direction. This is primarily due to its geometry: the depolarization field is parallel to the polarization, and changing the direction of polarization would be energetically very expensive. The (1$\bar{1}$0) DW has a small but noticeable Bloch-N{\'e}el character. The existence of the Bloch component of polarization at this DW is unique among the DWs considered. However, the Bloch and N{\'e}el components are too small to have a substantial effect on the electronic states. This is partially confirmed by the calculation of the local density of states (DOS) of this DW, which is very similar to the DOS of bulk GeTe.

\begin{figure}[h!]
\begin{center}
\includegraphics[width = 0.9\linewidth]{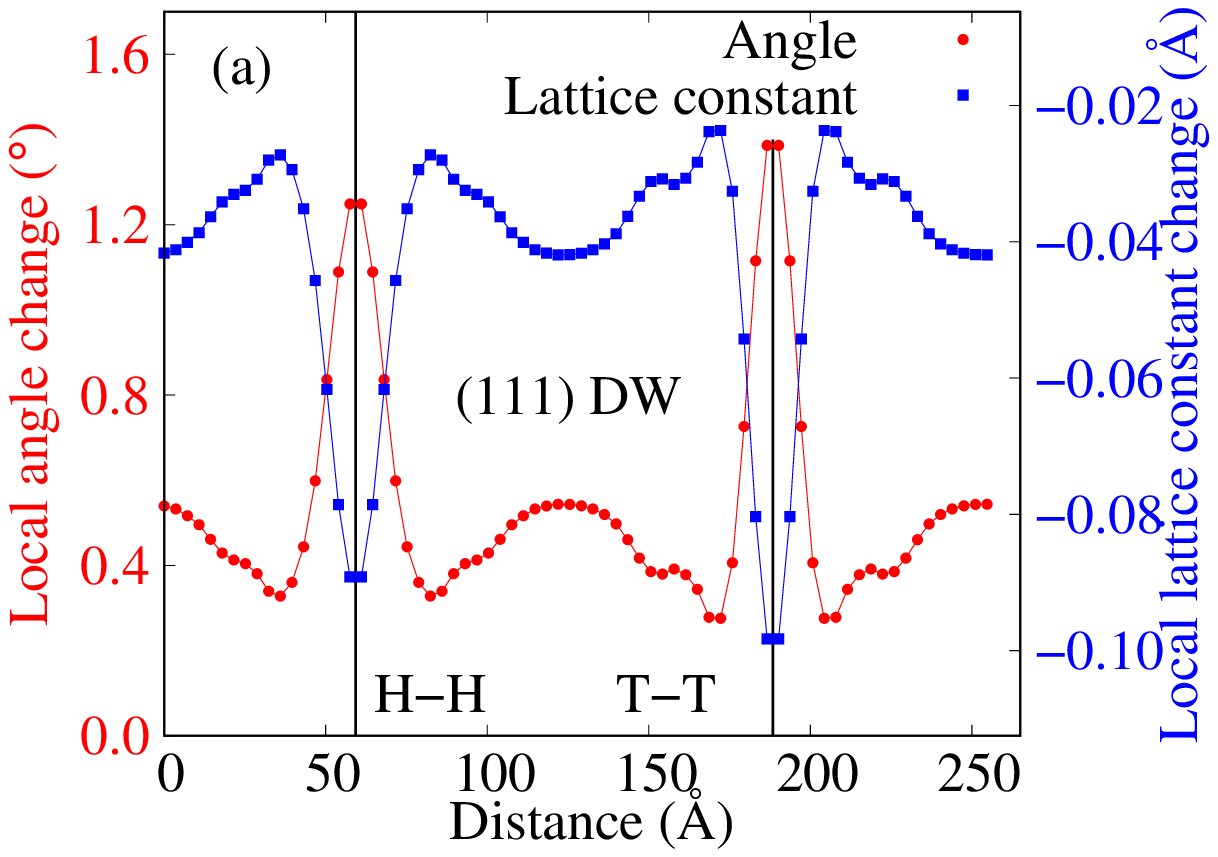}
\includegraphics[width = 0.9\linewidth]{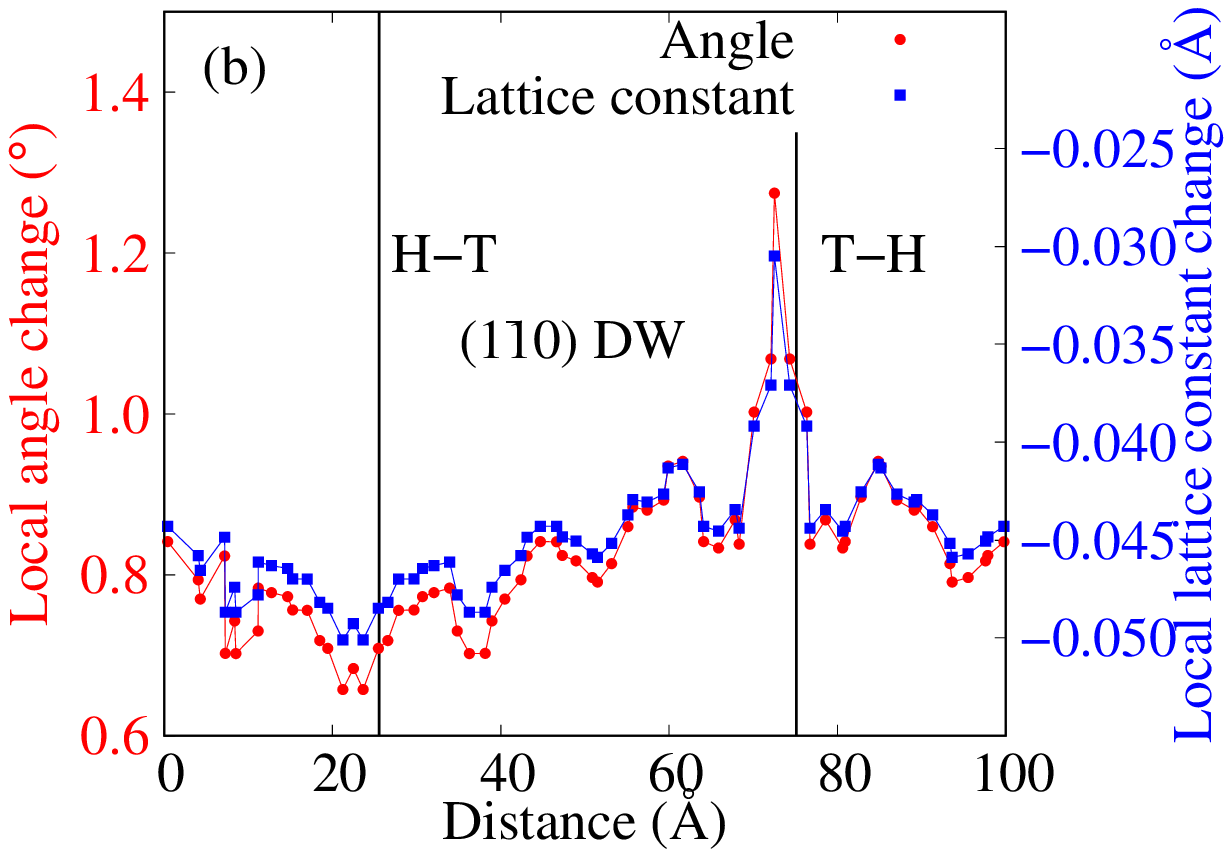}
\caption{Local lattice constant (blue line) and local rhombohedral angle (red line) for GeTe structures containing (a) (111) and (b) (1$\bar{1}$0) domain walls (DWs). DW boundaries are indicated by black vertical lines and labeled T-T for tail-to-tail, H-H for head-to-head, H-T for head-to-tail and T-H for tail-to-head DWs.}
\label{fig7}
\end{center}
\end{figure}

Fig.~\ref{fig7} illustrates local structural distortions for (111) and ($1\bar{1}0$) DWs. Both (111) DWs have comparatively large changes of the local angle and lattice constant, similarly to ($11\bar{1}$) DWs. They exhibit the same trend as single crystalline GeTe near the phase transition~\cite{main, newmain, Mi}: increasing angle and decreasing lattice constant and polarization closer to the DW. This is expected due to the pure Ising character of this DW and the fact that depolarizing field is collinear with polarization. Local structural distortions of  (1$\bar{1}$0) DW resemble numerical noise, since the structural changes along domains are smaller than the renormalization from the bulk values in the middle of domains. These effects as well as the differences in the polarization profiles and DW widths in (1$\bar{1}$0) DWs may come from a small domain size used in our calculations. Using larger supercells is computationally expensive and the properties of ($1\bar{1}0$) DW are not of significant immediate interest. 

\section{Thermal conductivity of GeTe with domain walls}

Large local changes of the lattice constant and angle in the vicinity of DWs driven by large polarization changes indicate the presence of strong strain-order parameter coupling. This mechanism is similar to acoustic-soft TO mode coupling in bulk GeTe~\cite{Ivana1}, and will likely reduce the lattice thermal conductivity of GeTe samples containing DWs in the direction perpendicular to the DW plane. Alternatively, we can view DWs as grain boundaries which would effectively scatter phonons \cite{DWThermCond1,DWThermCond2}. Domain size will also determine the strength of phonon scattering. 39$^{\circ}$ and 141$^{\circ}$ DWs may be more beneficial for the lattice thermal conductivity reduction due to the larger lattice orientation mismatch (caused by different orientations of the Brillouin zone in neighbouring domains) at the DW boundary.

We compute phonon scattering rates due to strain-order parameter coupling and lattice orientation mismatch caused by DWs as follows. First we define the scattering potential \cite{Hanus2018} with respect to structural deformations as:
\begin{align*}
V^{j}_{u}(\vec{k},x) = \hbar \Delta\omega ^{j}(\vec{k},x)=\hbar \omega ^{j}(\vec{k}) \gamma ^{j}_{u}(\vec{k}) \epsilon _{u}(x). \numberthis
\label{eq5}
\end{align*} 
Here $u$ represents one of the structural parameters: $u = a,\theta ,\tau$ ($a$ - lattice constant, $\theta$ - angle, $\tau$ - internal atomic displacement), $\omega ^{j}(\vec{k})$ is the frequency of the phonon mode with the wave vector $\vec{k}$ and the branch $j$, $\Delta \omega ^{j}(\vec{k},x)$ is the change of $\omega ^{j}(\vec{k})$ at position $x$ along the structure, and $\epsilon _{u}(x)$ is the relative change of the structural parameter $u$ at $x$ with respect to bulk. $\gamma ^{j}_{u}(\vec{k})$ is the generalized mode Gr{\"u}neisen parameter defined as:
\begin{align*}
\gamma ^{j}_{u}(\vec{k}) = -\frac{u}{\omega ^{j} (\vec{k})}\frac{\partial \omega ^{j} (\vec{k})}{\partial u},\numberthis
\end{align*}
which we also used to accurately describe the thermal expansion of GeTe \cite{Mi}.

To account for the lattice orientation mismatch at the domain wall boundaries in 39$^{\circ}$ and 141$^{\circ}$ ($11\bar{1}$) DWs, we consider the cases of phonon reflection and transmission that conserve the phonon momentum inside the DW plane \cite{AMM}. A phonon with certain Cartesian components of the momentum corresponds to different parts of the Brillouin zone depending on which side of the DW that phonon is. To preserve the momentum, the transmitting phonon needs to change its energy due to lattice orientation mismatch. The corresponding perturbation potential can then be defined as:
\begin{align*}
V^{j}_{m} (\vec{k},x) = \hbar \Delta \omega ^{j}(\vec{k}) \delta (x-d_{0}),
\numberthis
\label{eq6}
\end{align*}
where $\Delta \omega ^{j}(\vec{k})$ is the phonon frequency change for the phonon with frequency  $\omega ^{j}(\vec{k})$ whose momentum is conserved upon transmission to the other side of the DW, and $d_{0}$ is the position of the DW in the structure.

Using Fermi's golden rule, we define the scattering rate induced by the presence of a DW as \cite{Hanus2018}:
\begin{align*}
\Gamma (\vec{k}) = \frac{n}{v_{g}\hbar ^2}\frac{2k_{x}^2}{k^2}|g(2k_{x})|^2, \numberthis
\label{eq9}
\end{align*}
where $n=1/2L$ is the density of DWs, $2L$ is the domain size, $v_{g}$ is the group velocity in the direction of the DW vector, $k_{x}$ is the projection of the phonon wave vector in same direction and $|g(2k_{x})|$ is the Fourier transform of the scattering potential:
\begin{align*}
g(2k_{x}) = \int_{-L}^{L}V(x) e^{-i2k_{x}x} dx. \numberthis
\end{align*}

We have made the following approximations in the implementation of the outlined approach. The polarization change in each calculation is taken to be purely Ising, so there is no phonon scattering due to rotation of polarization. Gr{\"u}neisen parameters do not accurately quantify phonon frequency changes for large structural distortions. We do not account for the structural renormalization in the middle of domains. We assume that this effect arises due to finite size effects in our calculations and should be zero for domain sizes of $\sim$100 nm.
For ($1\bar{1}0$) DW, we do not account for the observed small changes in the lattice constant and angle. We do not take into account diffusive scattering at DWs, since this effect may not be important \cite{DWThermCond, DWThermCond1}. The results obtained with our model should represent a lower bound for the lattice thermal conductivity reduction due to DWs. 

We calculate the lattice thermal conductivity of GeTe with domain walls in the direction perpendicular to DW planes as:
\begin{align*}
\kappa _{L}= \frac{1}{NV}\sum_{\vec{k},j} c(\omega ^{j}(\vec{k}))(v^{j}(\vec{k}))^2/(\Gamma ^{j}_{anh}(\vec{k}) + \Gamma ^{j}_{DW}(\vec{k})),\numberthis
\end{align*}
where $c(\omega ^{j} (\vec{k}))$ is the specific heat capacity of the phonon mode with the wave vector $\vec{k}$ and the branch $j$, $v^{j}(\vec{k})$ is its group velocity, and $\Gamma ^{j} _{anh} (\vec{k})$ and $\Gamma ^{j} _{DW} (\vec{k})$ are the phonon scattering rates due to anharmonic processes and DWs, respectively. Here we used the constant relaxation time approximation for $\Gamma ^{j} _{anh} (\vec{k})$. We calculated this value at several different temperatures from our previous calculations of the lattice thermal conductivity of GeTe \citep{Ivana1}.

\begin{figure}[t]
\begin{center}
\includegraphics[width = 0.9\linewidth]{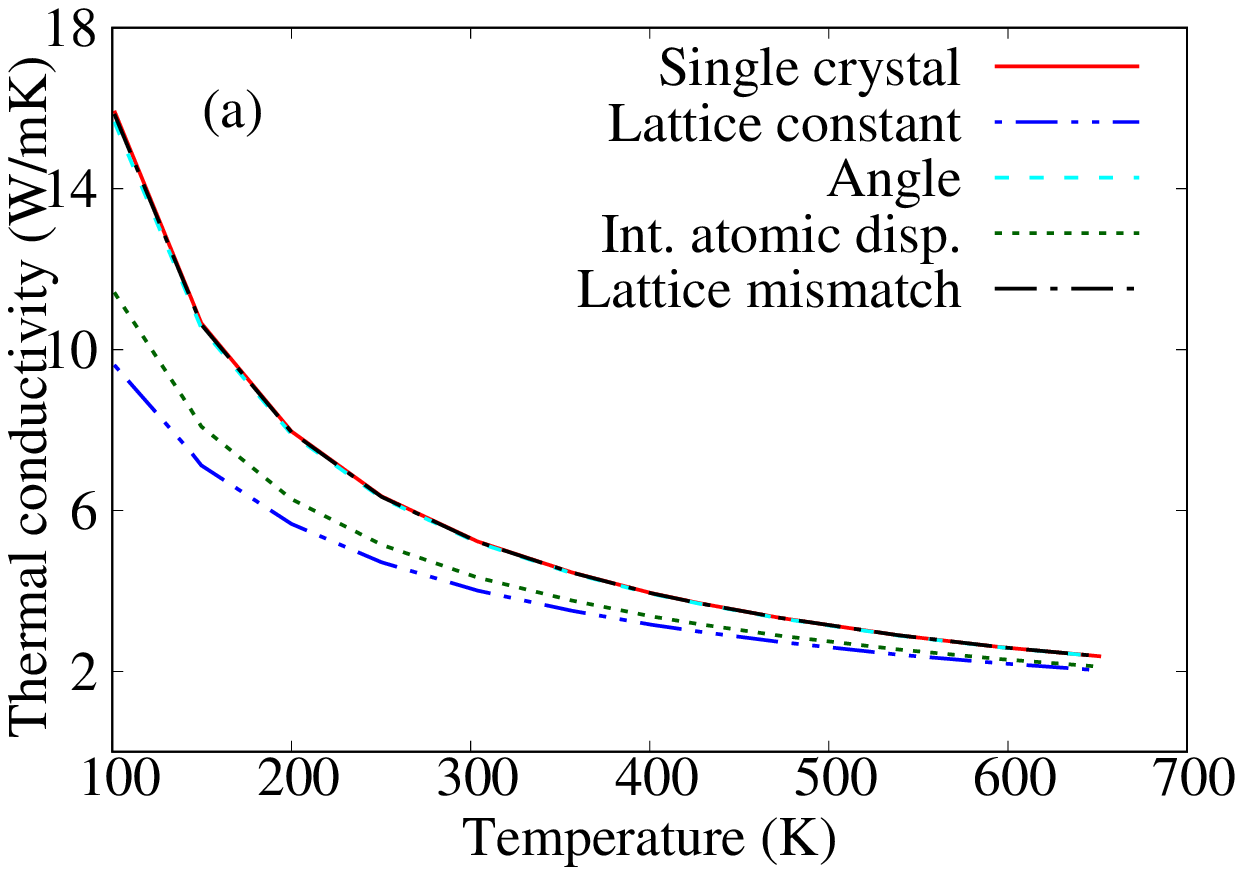}
\includegraphics[width = 0.9\linewidth]{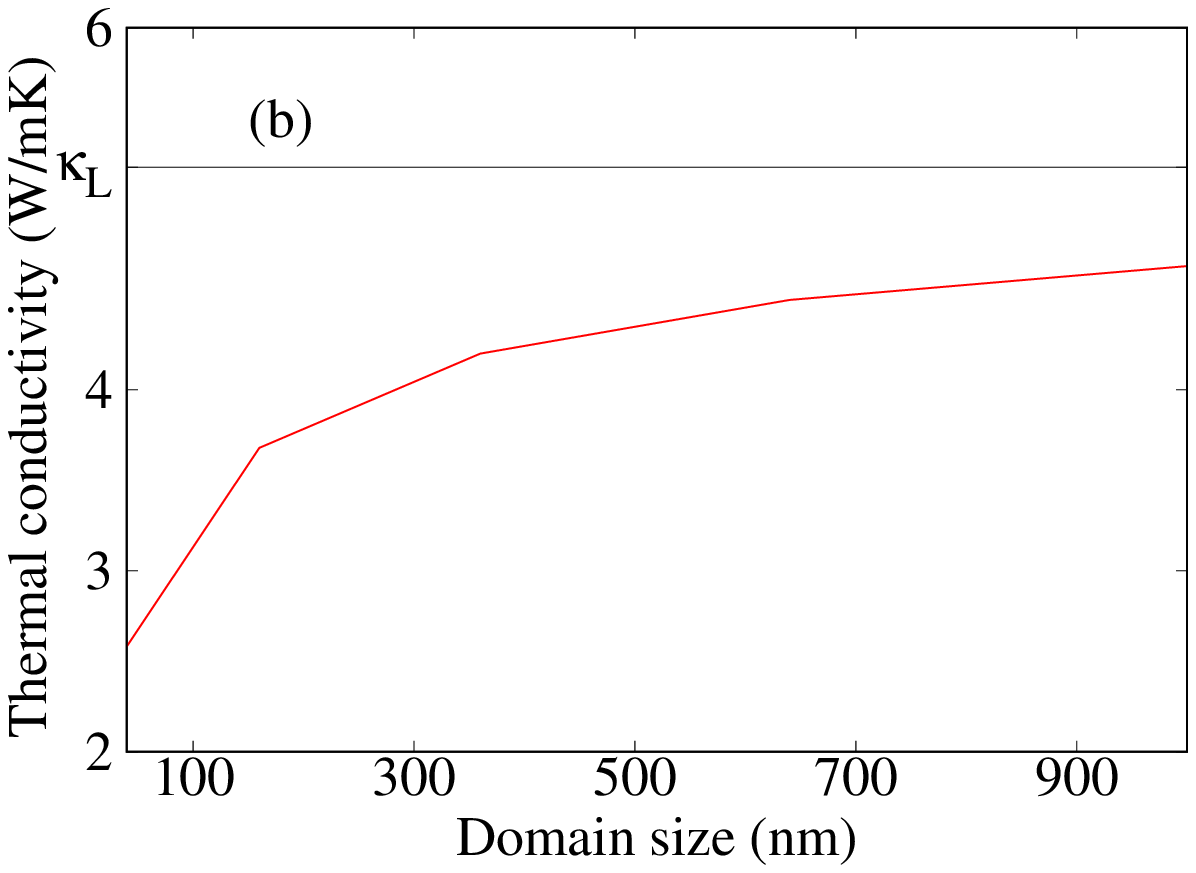}
\caption{(a) Lattice thermal conductivity of bulk GeTe and GeTe structure with 39$^{\circ}$ head-to-head (H-H) domain walls (DWs) and the domain size of 160 nm at 300 K, showing the contribution of each scattering mechanism due to the DWs. Contributions from lattice orientation mismatch and the change of angle to the conductivity reduction are negligible and hence these lines are on top of the single crystal result. (b)~Dependence of the lattice thermal conductivity of GeTe with 39$^{\circ}$ H-H DWs on the domain size at 300 K.}
\label{fig8}
\end{center}
\end{figure}

Fig.~\ref{fig8}(a) shows the contribution of each type of scattering to the lattice thermal conductivity reduction in GeTe structure with 39$^{\circ}$ H-H DWs and the domain size of 160 nm (approximately the value observed in experiment \citep{ActaMatGeTeDW}) with respect to bulk. We assume that the perturbation potential is the sum of the contributions from lattice orientation mismatch and local changes of the structural parameters. This allows us to check the individual contributions of each perturbation potential (see Eqs.~\ref{eq5} and ~\ref{eq6}). The largest contributions to $\kappa_L$ come from local changes of the lattice constant and internal atomic displacement near the DWs, thus confirming that strong strain-order parameter coupling at DWs indeed reduces the $\kappa_L$ of GeTe. Local angle distortions have a weak effect on $\kappa_L$ due to relatively small values of the generalized mode Gr{\"u}neisen parameters for the angle. The contribution of lattice orientation mismatch to the $\kappa_L$ reduction is also small since the domain size is much larger than the average phonon mean free path in GeTe. At higher temperatures, the difference between the $\kappa_L$ values for single crystal and domain structure becomes smaller since anharmonic processes become dominant. Fig.~\ref{fig8}(b) illustrates the effect of the domain size on the lattice thermal conductivity at room temperature. For large domain sizes, $\kappa_L$ tends towards the bulk value. For smaller domain sizes, there is a steep decline of the lattice thermal conductivity, driven by an increased density of local structural distortions. 

Table \ref{tb3} shows considerable reductions of the lattice thermal conductivity of GeTe structure with one particular type of considered DWs and the domain size of 160 nm with respect to bulk at 300 K. (111) DWs have large thermal resistance since they are the widest DWs considered and have sizeable structural distortions. Similarly, 180$^{\circ}$ ($11\bar{1}$) T-T DW has the largest thermal resistance due to large structural deformations in the vicinity of this relatively wide DW. The smallest reduction of $\kappa_L$ is obtained for ($1\bar{1}0$) DW, where local changes of the lattice constant and angle are taken to be zero. Consequently, the larger the amount of local distortions near the DW, the larger the $\kappa_L$ reduction. Our results clearly illustrate the potential of domain walls for substantially reducing the lattice thermal conductivity of GeTe.

\begin{table}[h]
\begin{center}
\begin{tabularx}{0.5\textwidth}{ p{2.2cm} | Y Y Y Y Y Y }
\hline \hline
   Charged DW    &  \multicolumn{2}{Y|}{39$^{\circ}$ (11$\bar{1}$)} &  \multicolumn{2}{Y|}{180$^{\circ}$ (11$\bar{1}$)} & \multicolumn{2}{Y}{180$^{\circ}$ (111)} \\ \hline
H-H DW &  \multicolumn{2}{Y|}{70\%} & \multicolumn{2}{Y|}{71\%} & \multicolumn{2}{Y}{ 65\%} \\ \hline
T-T DW & \multicolumn{2}{Y|}{79\%} & \multicolumn{2}{Y|}{53\%} & \multicolumn{2}{Y}{64\%} \\ \hline
 \hline
 \multicolumn{7}{@{}c@{}}{\begin{tabularx}{\dimexpr 0.5\textwidth-2\arrayrulewidth}[t] { p{2.19cm} | Y | Y }
  Neutral DW & 141$^{\circ}$ (11$\bar{1}$) & 180$^{\circ}$ (1$\bar{1}$0) \\
    \hline
    H-T DW & 75\% & 79\% \\
  \end{tabularx}} \\ \hline
  \hline
\end{tabularx}
\end{center}
\caption{Lattice thermal conductivity of GeTe structure with a particular type of domain walls (DWs) at 300 K and the DW density of 1/160 nm$^{-1}$, given as a percentage of the bulk value. H-H and T-T denote head-to-head and tail-to-tail DWs, respectively. 141$^{\circ}$ and (1$\bar{1}$0) DW only have head-to-tail DWs.}
\label{tb3}
\end{table}

\section{Conclusion}

In summary, we presented a first principles structural characterization of (11$\bar{1}$), (111) and (1$\bar{1}$0) domain walls in GeTe, which included calculations of domain wall energies and widths, local polarization and local structure distortions. ($111$) domain walls exhibit the Ising character of polarization change, while all other domain walls show mixed Ising-N{\'e}el character. Large local structure distortions and strong strain-order parameter coupling are present at most of the domain walls investigated. We have shown that the lattice thermal conductivity of GeTe can be substantially lowered by these domain walls, particularly by those with large structural changes and large widths. At high domain wall densities, phonon scattering from strain fields becomes dominant, and lattice thermal conductivity can be dramatically suppressed. Domain engineering could thus be used to optimize the thermoelectric performance of GeTe.

\section{Acknowledgements}

We thank Mohammad Noor-A-Alam, Raymond MacQuaid and Marty Gregg for useful discussions. This work is supported by Science Foundation Ireland under grant numbers 15/IA/3160 and 13/RC/2077. The later grant is co-funded under the European Regional Development Fund. We acknowledge the Irish Centre for High-End Computing (ICHEC) for the provision of computational facilities.

\bibliography{main}{}

\begin{thebibliography}{57}%
\makeatletter
\providecommand \@ifxundefined [1]{%
 \@ifx{#1\undefined}
}%
\providecommand \@ifnum [1]{%
 \ifnum #1\expandafter \@firstoftwo
 \else \expandafter \@secondoftwo
 \fi
}%
\providecommand \@ifx [1]{%
 \ifx #1\expandafter \@firstoftwo
 \else \expandafter \@secondoftwo
 \fi
}%
\providecommand \natexlab [1]{#1}%
\providecommand \enquote  [1]{``#1''}%
\providecommand \bibnamefont  [1]{#1}%
\providecommand \bibfnamefont [1]{#1}%
\providecommand \citenamefont [1]{#1}%
\providecommand \href@noop [0]{\@secondoftwo}%
\providecommand \href [0]{\begingroup \@sanitize@url \@href}%
\providecommand \@href[1]{\@@startlink{#1}\@@href}%
\providecommand \@@href[1]{\endgroup#1\@@endlink}%
\providecommand \@sanitize@url [0]{\catcode `\\12\catcode `\$12\catcode
  `\&12\catcode `\#12\catcode `\^12\catcode `\_12\catcode `\%12\relax}%
\providecommand \@@startlink[1]{}%
\providecommand \@@endlink[0]{}%
\providecommand \url  [0]{\begingroup\@sanitize@url \@url }%
\providecommand \@url [1]{\endgroup\@href {#1}{\urlprefix }}%
\providecommand \urlprefix  [0]{URL }%
\providecommand \Eprint [0]{\href }%
\providecommand \doibase [0]{http://dx.doi.org/}%
\providecommand \selectlanguage [0]{\@gobble}%
\providecommand \bibinfo  [0]{\@secondoftwo}%
\providecommand \bibfield  [0]{\@secondoftwo}%
\providecommand \translation [1]{[#1]}%
\providecommand \BibitemOpen [0]{}%
\providecommand \bibitemStop [0]{}%
\providecommand \bibitemNoStop [0]{.\EOS\space}%
\providecommand \EOS [0]{\spacefactor3000\relax}%
\providecommand \BibitemShut  [1]{\csname bibitem#1\endcsname}%
\let\auto@bib@innerbib\@empty
\bibitem [{\citenamefont {Levin}\ \emph {et~al.}(2013)\citenamefont {Levin},
  \citenamefont {Besser},\ and\ \citenamefont {Hanus}}]{GeTe2}%
  \BibitemOpen
  \bibfield  {author} {\bibinfo {author} {\bibfnamefont {E.~M.}\ \bibnamefont
  {Levin}}, \bibinfo {author} {\bibfnamefont {M.~F.}\ \bibnamefont {Besser}}, \
  and\ \bibinfo {author} {\bibfnamefont {R.}~\bibnamefont {Hanus}},\ }\href
  {\doibase 10.1063/1.4819222} {\bibfield  {journal} {\bibinfo  {journal} {J.
  Appl. Phys.}\ }\textbf {\bibinfo {volume} {114}},\ \bibinfo {pages} {083713}
  (\bibinfo {year} {2013})}\BibitemShut {NoStop}%
\bibitem [{\citenamefont {Wu}\ \emph {et~al.}(2014)\citenamefont {Wu},
  \citenamefont {Zhao}, \citenamefont {Hao}, \citenamefont {Jiang},
  \citenamefont {Zheng}, \citenamefont {Doak}, \citenamefont {Wu},
  \citenamefont {Chi}, \citenamefont {Gelbstein}, \citenamefont {Uher},
  \citenamefont {Wolverton}, \citenamefont {Kanatzidis},\ and\ \citenamefont
  {He}}]{gete-jacs}%
  \BibitemOpen
  \bibfield  {author} {\bibinfo {author} {\bibfnamefont {D.}~\bibnamefont
  {Wu}}, \bibinfo {author} {\bibfnamefont {L.-D.}\ \bibnamefont {Zhao}},
  \bibinfo {author} {\bibfnamefont {S.}~\bibnamefont {Hao}}, \bibinfo {author}
  {\bibfnamefont {Q.}~\bibnamefont {Jiang}}, \bibinfo {author} {\bibfnamefont
  {F.}~\bibnamefont {Zheng}}, \bibinfo {author} {\bibfnamefont {J.~W.}\
  \bibnamefont {Doak}}, \bibinfo {author} {\bibfnamefont {H.}~\bibnamefont
  {Wu}}, \bibinfo {author} {\bibfnamefont {H.}~\bibnamefont {Chi}}, \bibinfo
  {author} {\bibfnamefont {Y.}~\bibnamefont {Gelbstein}}, \bibinfo {author}
  {\bibfnamefont {C.}~\bibnamefont {Uher}}, \bibinfo {author} {\bibfnamefont
  {C.}~\bibnamefont {Wolverton}}, \bibinfo {author} {\bibfnamefont
  {M.}~\bibnamefont {Kanatzidis}}, \ and\ \bibinfo {author} {\bibfnamefont
  {J.}~\bibnamefont {He}},\ }\href {https://doi.org/10.1021/ja504896a}
  {\bibfield  {journal} {\bibinfo  {journal} {J. Am. Chem. Soc.}\ }\textbf
  {\bibinfo {volume} {136}},\ \bibinfo {pages} {11412} (\bibinfo {year}
  {2014})}\BibitemShut {NoStop}%
\bibitem [{\citenamefont {Madar}\ \emph {et~al.}(2016)\citenamefont {Madar},
  \citenamefont {Givon}, \citenamefont {Mogilyansky},\ and\ \citenamefont
  {Gelbstein}}]{yaniv-gete-jap-16}%
  \BibitemOpen
  \bibfield  {author} {\bibinfo {author} {\bibfnamefont {N.}~\bibnamefont
  {Madar}}, \bibinfo {author} {\bibfnamefont {T.}~\bibnamefont {Givon}},
  \bibinfo {author} {\bibfnamefont {D.}~\bibnamefont {Mogilyansky}}, \ and\
  \bibinfo {author} {\bibfnamefont {Y.}~\bibnamefont {Gelbstein}},\ }\href
  {\doibase 10.1063/1.4958973} {\bibfield  {journal} {\bibinfo  {journal} {J.
  Appl. Phys.}\ }\textbf {\bibinfo {volume} {120}},\ \bibinfo {pages} {035102}
  (\bibinfo {year} {2016})}\BibitemShut {NoStop}%
\bibitem [{\citenamefont {Perumal}\ \emph {et~al.}(2016)\citenamefont
  {Perumal}, \citenamefont {Roychowdhury},\ and\ \citenamefont
  {Biswas}}]{biswas-gete-rev}%
  \BibitemOpen
  \bibfield  {author} {\bibinfo {author} {\bibfnamefont {S.}~\bibnamefont
  {Perumal}}, \bibinfo {author} {\bibfnamefont {S.}~\bibnamefont
  {Roychowdhury}}, \ and\ \bibinfo {author} {\bibfnamefont {K.}~\bibnamefont
  {Biswas}},\ }\href {http://dx.doi.org/10.1039/C6TC02501C} {\bibfield
  {journal} {\bibinfo  {journal} {J. Mater. Chem. C}\ }\textbf {\bibinfo
  {volume} {4}},\ \bibinfo {pages} {7520} (\bibinfo {year} {2016})}\BibitemShut
  {NoStop}%
\bibitem [{\citenamefont {Li}\ \emph {et~al.}(2017)\citenamefont {Li},
  \citenamefont {Chen}, \citenamefont {Zhang}, \citenamefont {Sun},
  \citenamefont {Yang},\ and\ \citenamefont {Pei}}]{GeTe3}%
  \BibitemOpen
  \bibfield  {author} {\bibinfo {author} {\bibfnamefont {J.}~\bibnamefont
  {Li}}, \bibinfo {author} {\bibfnamefont {Z.}~\bibnamefont {Chen}}, \bibinfo
  {author} {\bibfnamefont {X.}~\bibnamefont {Zhang}}, \bibinfo {author}
  {\bibfnamefont {Y.}~\bibnamefont {Sun}}, \bibinfo {author} {\bibfnamefont
  {J.}~\bibnamefont {Yang}}, \ and\ \bibinfo {author} {\bibfnamefont
  {Y.}~\bibnamefont {Pei}},\ }\href {http://dx.doi.org/10.1038/am.2017.8}
  {\bibfield  {journal} {\bibinfo  {journal} {NPG Asia Mater.}\ }\textbf
  {\bibinfo {volume} {9}},\ \bibinfo {pages} {e353} (\bibinfo {year}
  {2017})}\BibitemShut {NoStop}%
\bibitem [{\citenamefont {Zheng}\ \emph {et~al.}(2018)\citenamefont {Zheng},
  \citenamefont {Su}, \citenamefont {Deng}, \citenamefont {Stoumpos},
  \citenamefont {Xie}, \citenamefont {Liu}, \citenamefont {Yan}, \citenamefont
  {Hao}, \citenamefont {Uher}, \citenamefont {Wolverton}, \citenamefont
  {Kanatzidis},\ and\ \citenamefont {Tang}}]{GeMnTe}%
  \BibitemOpen
  \bibfield  {author} {\bibinfo {author} {\bibfnamefont {Z.}~\bibnamefont
  {Zheng}}, \bibinfo {author} {\bibfnamefont {X.}~\bibnamefont {Su}}, \bibinfo
  {author} {\bibfnamefont {R.}~\bibnamefont {Deng}}, \bibinfo {author}
  {\bibfnamefont {C.}~\bibnamefont {Stoumpos}}, \bibinfo {author}
  {\bibfnamefont {H.}~\bibnamefont {Xie}}, \bibinfo {author} {\bibfnamefont
  {W.}~\bibnamefont {Liu}}, \bibinfo {author} {\bibfnamefont {Y.}~\bibnamefont
  {Yan}}, \bibinfo {author} {\bibfnamefont {S.}~\bibnamefont {Hao}}, \bibinfo
  {author} {\bibfnamefont {C.}~\bibnamefont {Uher}}, \bibinfo {author}
  {\bibfnamefont {C.}~\bibnamefont {Wolverton}}, \bibinfo {author}
  {\bibfnamefont {M.~G.}\ \bibnamefont {Kanatzidis}}, \ and\ \bibinfo {author}
  {\bibfnamefont {X.}~\bibnamefont {Tang}},\ }\href {\doibase
  10.1021/jacs.7b13611} {\bibfield  {journal} {\bibinfo  {journal} {J. Am.
  Chem. Soc.}\ }\textbf {\bibinfo {volume} {140}},\ \bibinfo {pages} {2673}
  (\bibinfo {year} {2018})}\BibitemShut {NoStop}%
\bibitem [{\citenamefont {Roychowdhury}\ \emph {et~al.}(2018)\citenamefont
  {Roychowdhury}, \citenamefont {Samanta}, \citenamefont {Perumal},\ and\
  \citenamefont {Biswas}}]{GeTe1}%
  \BibitemOpen
  \bibfield  {author} {\bibinfo {author} {\bibfnamefont {S.}~\bibnamefont
  {Roychowdhury}}, \bibinfo {author} {\bibfnamefont {M.}~\bibnamefont
  {Samanta}}, \bibinfo {author} {\bibfnamefont {S.}~\bibnamefont {Perumal}}, \
  and\ \bibinfo {author} {\bibfnamefont {K.}~\bibnamefont {Biswas}},\ }\href
  {\doibase 10.1021/acs.chemmater.8b02676} {\bibfield  {journal} {\bibinfo
  {journal} {Chem. Mater.}\ }\textbf {\bibinfo {volume} {30}},\ \bibinfo
  {pages} {5799} (\bibinfo {year} {2018})}\BibitemShut {NoStop}%
\bibitem [{\citenamefont {Liu}\ \emph {et~al.}(2018)\citenamefont {Liu},
  \citenamefont {Sun}, \citenamefont {Mao}, \citenamefont {Zhu}, \citenamefont
  {Ren}, \citenamefont {Zhou}, \citenamefont {Wang}, \citenamefont {Singh},
  \citenamefont {Sui}, \citenamefont {Chu},\ and\ \citenamefont
  {Ren}}]{GeTePNAS}%
  \BibitemOpen
  \bibfield  {author} {\bibinfo {author} {\bibfnamefont {Z.}~\bibnamefont
  {Liu}}, \bibinfo {author} {\bibfnamefont {J.}~\bibnamefont {Sun}}, \bibinfo
  {author} {\bibfnamefont {J.}~\bibnamefont {Mao}}, \bibinfo {author}
  {\bibfnamefont {H.}~\bibnamefont {Zhu}}, \bibinfo {author} {\bibfnamefont
  {W.}~\bibnamefont {Ren}}, \bibinfo {author} {\bibfnamefont {J.}~\bibnamefont
  {Zhou}}, \bibinfo {author} {\bibfnamefont {Z.}~\bibnamefont {Wang}}, \bibinfo
  {author} {\bibfnamefont {D.~J.}\ \bibnamefont {Singh}}, \bibinfo {author}
  {\bibfnamefont {J.}~\bibnamefont {Sui}}, \bibinfo {author} {\bibfnamefont
  {C.~W.}\ \bibnamefont {Chu}}, \ and\ \bibinfo {author} {\bibfnamefont
  {Z.}~\bibnamefont {Ren}},\ }\href {https://doi.org/10.1073/pnas.1802020115}
  {\bibfield  {journal} {\bibinfo  {journal} {Proc. Natl. Acad. Sci. U.S.A}\
  }\textbf {\bibinfo {volume} {115}},\ \bibinfo {pages} {5332} (\bibinfo {year}
  {2018})}\BibitemShut {NoStop}%
\bibitem [{\citenamefont {Li}\ \emph {et~al.}(2018)\citenamefont {Li},
  \citenamefont {Zhang}, \citenamefont {Chen}, \citenamefont {Lin},
  \citenamefont {Li}, \citenamefont {Shen}, \citenamefont {Witting},
  \citenamefont {Faghaninia}, \citenamefont {Chen}, \citenamefont {Jain},
  \citenamefont {Chen}, \citenamefont {Snyder},\ and\ \citenamefont
  {Pei}}]{Li2018}%
  \BibitemOpen
  \bibfield  {author} {\bibinfo {author} {\bibfnamefont {J.}~\bibnamefont
  {Li}}, \bibinfo {author} {\bibfnamefont {X.}~\bibnamefont {Zhang}}, \bibinfo
  {author} {\bibfnamefont {Z.}~\bibnamefont {Chen}}, \bibinfo {author}
  {\bibfnamefont {S.}~\bibnamefont {Lin}}, \bibinfo {author} {\bibfnamefont
  {W.}~\bibnamefont {Li}}, \bibinfo {author} {\bibfnamefont {J.}~\bibnamefont
  {Shen}}, \bibinfo {author} {\bibfnamefont {I.~T.}\ \bibnamefont {Witting}},
  \bibinfo {author} {\bibfnamefont {A.}~\bibnamefont {Faghaninia}}, \bibinfo
  {author} {\bibfnamefont {Y.}~\bibnamefont {Chen}}, \bibinfo {author}
  {\bibfnamefont {A.}~\bibnamefont {Jain}}, \bibinfo {author} {\bibfnamefont
  {L.}~\bibnamefont {Chen}}, \bibinfo {author} {\bibfnamefont {G.~J.}\
  \bibnamefont {Snyder}}, \ and\ \bibinfo {author} {\bibfnamefont
  {Y.}~\bibnamefont {Pei}},\ }\href {\doibase
  https://doi.org/10.1016/j.joule.2018.02.016} {\bibfield  {journal} {\bibinfo
  {journal} {Joule}\ }\textbf {\bibinfo {volume} {2}},\ \bibinfo {pages} {976 }
  (\bibinfo {year} {2018})}\BibitemShut {NoStop}%
\bibitem [{\citenamefont {Zhang}\ \emph {et~al.}(2018)\citenamefont {Zhang},
  \citenamefont {Li}, \citenamefont {Wang}, \citenamefont {Chen}, \citenamefont
  {Mao}, \citenamefont {Chen},\ and\ \citenamefont {Pei}}]{Zhang2018}%
  \BibitemOpen
  \bibfield  {author} {\bibinfo {author} {\bibfnamefont {X.}~\bibnamefont
  {Zhang}}, \bibinfo {author} {\bibfnamefont {J.}~\bibnamefont {Li}}, \bibinfo
  {author} {\bibfnamefont {X.}~\bibnamefont {Wang}}, \bibinfo {author}
  {\bibfnamefont {Z.}~\bibnamefont {Chen}}, \bibinfo {author} {\bibfnamefont
  {J.}~\bibnamefont {Mao}}, \bibinfo {author} {\bibfnamefont {Y.}~\bibnamefont
  {Chen}}, \ and\ \bibinfo {author} {\bibfnamefont {Y.}~\bibnamefont {Pei}},\
  }\href {\doibase 10.1021/jacs.8b09375} {\bibfield  {journal} {\bibinfo
  {journal} {J. Am. Chem. Soc.}\ }\textbf {\bibinfo {volume} {140}},\ \bibinfo
  {pages} {15883} (\bibinfo {year} {2018})}\BibitemShut {NoStop}%
\bibitem [{\citenamefont {Dong}\ \emph {et~al.}(2019)\citenamefont {Dong},
  \citenamefont {Sun}, \citenamefont {Tang}, \citenamefont {Pei}, \citenamefont
  {Zhuang}, \citenamefont {Hu}, \citenamefont {Zhang}, \citenamefont {Pan},\
  and\ \citenamefont {Li}}]{Dong2019}%
  \BibitemOpen
  \bibfield  {author} {\bibinfo {author} {\bibfnamefont {J.}~\bibnamefont
  {Dong}}, \bibinfo {author} {\bibfnamefont {F.-H.}\ \bibnamefont {Sun}},
  \bibinfo {author} {\bibfnamefont {H.}~\bibnamefont {Tang}}, \bibinfo {author}
  {\bibfnamefont {J.}~\bibnamefont {Pei}}, \bibinfo {author} {\bibfnamefont
  {H.-L.}\ \bibnamefont {Zhuang}}, \bibinfo {author} {\bibfnamefont {H.-H.}\
  \bibnamefont {Hu}}, \bibinfo {author} {\bibfnamefont {B.-P.}\ \bibnamefont
  {Zhang}}, \bibinfo {author} {\bibfnamefont {Y.}~\bibnamefont {Pan}}, \ and\
  \bibinfo {author} {\bibfnamefont {J.-F.}\ \bibnamefont {Li}},\ }\href
  {\doibase 10.1039/C9EE00317G} {\bibfield  {journal} {\bibinfo  {journal}
  {Energy Environ. Sci.}\ }\textbf {\bibinfo {volume} {12}},\ \bibinfo {pages}
  {1396} (\bibinfo {year} {2019})}\BibitemShut {NoStop}%
\bibitem [{\citenamefont {Wu}\ \emph {et~al.}(2019)\citenamefont {Wu},
  \citenamefont {Xie}, \citenamefont {Xu},\ and\ \citenamefont {He}}]{Wu2019}%
  \BibitemOpen
  \bibfield  {author} {\bibinfo {author} {\bibfnamefont {D.}~\bibnamefont
  {Wu}}, \bibinfo {author} {\bibfnamefont {L.}~\bibnamefont {Xie}}, \bibinfo
  {author} {\bibfnamefont {X.}~\bibnamefont {Xu}}, \ and\ \bibinfo {author}
  {\bibfnamefont {J.}~\bibnamefont {He}},\ }\href {\doibase
  10.1002/adfm.201806613} {\bibfield  {journal} {\bibinfo  {journal} {Adv.
  Funct. Mater.}\ }\textbf {\bibinfo {volume} {29}},\ \bibinfo {pages}
  {1806613} (\bibinfo {year} {2019})}\BibitemShut {NoStop}%
\bibitem [{\citenamefont {Hong}\ \emph {et~al.}(2019)\citenamefont {Hong},
  \citenamefont {Zou},\ and\ \citenamefont {Chen}}]{Hong2019}%
  \BibitemOpen
  \bibfield  {author} {\bibinfo {author} {\bibfnamefont {M.}~\bibnamefont
  {Hong}}, \bibinfo {author} {\bibfnamefont {J.}~\bibnamefont {Zou}}, \ and\
  \bibinfo {author} {\bibfnamefont {Z.-G.}\ \bibnamefont {Chen}},\ }\href
  {\doibase 10.1002/adma.201807071} {\bibfield  {journal} {\bibinfo  {journal}
  {Adv. Mater.}\ }\textbf {\bibinfo {volume} {31}},\ \bibinfo {pages} {1807071}
  (\bibinfo {year} {2019})}\BibitemShut {NoStop}%
\bibitem [{\citenamefont {Murphy}\ \emph {et~al.}(2017)\citenamefont {Murphy},
  \citenamefont {Murray}, \citenamefont {Fahy},\ and\ \citenamefont
  {Savi\ifmmode~\acute{c}\else \'{c}\fi{}}}]{Ivana1}%
  \BibitemOpen
  \bibfield  {author} {\bibinfo {author} {\bibfnamefont {R.~M.}\ \bibnamefont
  {Murphy}}, \bibinfo {author} {\bibfnamefont {{\'E}.~D.}\ \bibnamefont
  {Murray}}, \bibinfo {author} {\bibfnamefont {S.}~\bibnamefont {Fahy}}, \ and\
  \bibinfo {author} {\bibfnamefont {I.}~\bibnamefont
  {Savi\ifmmode~\acute{c}\else \'{c}\fi{}}},\ }\href {\doibase
  10.1103/PhysRevB.95.144302} {\bibfield  {journal} {\bibinfo  {journal} {Phys.
  Rev. B}\ }\textbf {\bibinfo {volume} {95}},\ \bibinfo {pages} {144302}
  (\bibinfo {year} {2017})}\BibitemShut {NoStop}%
\bibitem [{\citenamefont {Campi}\ \emph {et~al.}(2017)\citenamefont {Campi},
  \citenamefont {Paulatto}, \citenamefont {Fugallo}, \citenamefont {Mauri},\
  and\ \citenamefont {Bernasconi}}]{CampiGeTe}%
  \BibitemOpen
  \bibfield  {author} {\bibinfo {author} {\bibfnamefont {D.}~\bibnamefont
  {Campi}}, \bibinfo {author} {\bibfnamefont {L.}~\bibnamefont {Paulatto}},
  \bibinfo {author} {\bibfnamefont {G.}~\bibnamefont {Fugallo}}, \bibinfo
  {author} {\bibfnamefont {F.}~\bibnamefont {Mauri}}, \ and\ \bibinfo {author}
  {\bibfnamefont {M.}~\bibnamefont {Bernasconi}},\ }\href {\doibase
  10.1103/PhysRevB.95.024311} {\bibfield  {journal} {\bibinfo  {journal} {Phys.
  Rev. B}\ }\textbf {\bibinfo {volume} {95}},\ \bibinfo {pages} {024311}
  (\bibinfo {year} {2017})}\BibitemShut {NoStop}%
\bibitem [{\citenamefont {Vermeulen}\ \emph {et~al.}(2016)\citenamefont
  {Vermeulen}, \citenamefont {Kumar}, \citenamefont {ten Brink}, \citenamefont
  {Blake},\ and\ \citenamefont {Kooi}}]{DWGeTe1}%
  \BibitemOpen
  \bibfield  {author} {\bibinfo {author} {\bibfnamefont {P.~A.}\ \bibnamefont
  {Vermeulen}}, \bibinfo {author} {\bibfnamefont {A.}~\bibnamefont {Kumar}},
  \bibinfo {author} {\bibfnamefont {G.~H.}\ \bibnamefont {ten Brink}}, \bibinfo
  {author} {\bibfnamefont {G.~R.}\ \bibnamefont {Blake}}, \ and\ \bibinfo
  {author} {\bibfnamefont {B.~J.}\ \bibnamefont {Kooi}},\ }\href {\doibase
  10.1021/acs.cgd.6b00960} {\bibfield  {journal} {\bibinfo  {journal} {Cryst.
  Growth Des.}\ }\textbf {\bibinfo {volume} {16}},\ \bibinfo {pages} {5915}
  (\bibinfo {year} {2016})}\BibitemShut {NoStop}%
\bibitem [{\citenamefont {Nukala}\ \emph {et~al.}(2017)\citenamefont {Nukala},
  \citenamefont {Ren}, \citenamefont {Agarwal}, \citenamefont {Berger},
  \citenamefont {Liu}, \citenamefont {Johnson},\ and\ \citenamefont
  {Agarwal}}]{DWGeTe2}%
  \BibitemOpen
  \bibfield  {author} {\bibinfo {author} {\bibfnamefont {P.}~\bibnamefont
  {Nukala}}, \bibinfo {author} {\bibfnamefont {M.}~\bibnamefont {Ren}},
  \bibinfo {author} {\bibfnamefont {R.}~\bibnamefont {Agarwal}}, \bibinfo
  {author} {\bibfnamefont {J.}~\bibnamefont {Berger}}, \bibinfo {author}
  {\bibfnamefont {G.}~\bibnamefont {Liu}}, \bibinfo {author} {\bibfnamefont
  {A.~T.~C.}\ \bibnamefont {Johnson}}, \ and\ \bibinfo {author} {\bibfnamefont
  {R.}~\bibnamefont {Agarwal}},\ }\href {http://dx.doi.org/10.1038/ncomms15033}
  {\bibfield  {journal} {\bibinfo  {journal} {Nat. Commun.}\ }\textbf {\bibinfo
  {volume} {8}},\ \bibinfo {pages} {15033} (\bibinfo {year}
  {2017})}\BibitemShut {NoStop}%
\bibitem [{\citenamefont {Rinaldi}\ \emph {et~al.}(2018)\citenamefont
  {Rinaldi}, \citenamefont {Varotto}, \citenamefont {Asa}, \citenamefont
  {Slawi{\'n}ska}, \citenamefont {Fujii}, \citenamefont {Vinai}, \citenamefont
  {Cecchi}, \citenamefont {Di~Sante}, \citenamefont {Calarco}, \citenamefont
  {Vobornik}, \citenamefont {Panaccione}, \citenamefont {Picozzi},\ and\
  \citenamefont {Bertacco}}]{RashbaGeTe}%
  \BibitemOpen
  \bibfield  {author} {\bibinfo {author} {\bibfnamefont {C.}~\bibnamefont
  {Rinaldi}}, \bibinfo {author} {\bibfnamefont {S.}~\bibnamefont {Varotto}},
  \bibinfo {author} {\bibfnamefont {M.}~\bibnamefont {Asa}}, \bibinfo {author}
  {\bibfnamefont {J.}~\bibnamefont {Slawi{\'n}ska}}, \bibinfo {author}
  {\bibfnamefont {J.}~\bibnamefont {Fujii}}, \bibinfo {author} {\bibfnamefont
  {G.}~\bibnamefont {Vinai}}, \bibinfo {author} {\bibfnamefont
  {S.}~\bibnamefont {Cecchi}}, \bibinfo {author} {\bibfnamefont
  {D.}~\bibnamefont {Di~Sante}}, \bibinfo {author} {\bibfnamefont
  {R.}~\bibnamefont {Calarco}}, \bibinfo {author} {\bibfnamefont
  {I.}~\bibnamefont {Vobornik}}, \bibinfo {author} {\bibfnamefont
  {G.}~\bibnamefont {Panaccione}}, \bibinfo {author} {\bibfnamefont
  {S.}~\bibnamefont {Picozzi}}, \ and\ \bibinfo {author} {\bibfnamefont
  {R.}~\bibnamefont {Bertacco}},\ }\href {\doibase
  10.1021/acs.nanolett.7b04829} {\bibfield  {journal} {\bibinfo  {journal}
  {Nano Lett.}\ }\textbf {\bibinfo {volume} {18}},\ \bibinfo {pages} {2751}
  (\bibinfo {year} {2018})}\BibitemShut {NoStop}%
\bibitem [{\citenamefont {Lee}\ \emph {et~al.}(2015)\citenamefont {Lee},
  \citenamefont {Kim}, \citenamefont {Cho}, \citenamefont {Oh}, \citenamefont
  {Min}, \citenamefont {Park},\ and\ \citenamefont {Lee}}]{ActaMatGeTeDW}%
  \BibitemOpen
  \bibfield  {author} {\bibinfo {author} {\bibfnamefont {H.~S.}\ \bibnamefont
  {Lee}}, \bibinfo {author} {\bibfnamefont {B.-S.}\ \bibnamefont {Kim}},
  \bibinfo {author} {\bibfnamefont {C.-W.}\ \bibnamefont {Cho}}, \bibinfo
  {author} {\bibfnamefont {M.-W.}\ \bibnamefont {Oh}}, \bibinfo {author}
  {\bibfnamefont {B.-K.}\ \bibnamefont {Min}}, \bibinfo {author} {\bibfnamefont
  {S.-D.}\ \bibnamefont {Park}}, \ and\ \bibinfo {author} {\bibfnamefont
  {H.-W.}\ \bibnamefont {Lee}},\ }\href {\doibase
  https://doi.org/10.1016/j.actamat.2015.03.015} {\bibfield  {journal}
  {\bibinfo  {journal} {Acta Materialia}\ }\textbf {\bibinfo {volume} {91}},\
  \bibinfo {pages} {83 } (\bibinfo {year} {2015})}\BibitemShut {NoStop}%
\bibitem [{\citenamefont {Snykers}\ \emph {et~al.}(1972)\citenamefont
  {Snykers}, \citenamefont {Delavignette},\ and\ \citenamefont
  {Amelinckx}}]{SNYKERS1972}%
  \BibitemOpen
  \bibfield  {author} {\bibinfo {author} {\bibfnamefont {M.}~\bibnamefont
  {Snykers}}, \bibinfo {author} {\bibfnamefont {P.}~\bibnamefont
  {Delavignette}}, \ and\ \bibinfo {author} {\bibfnamefont {S.}~\bibnamefont
  {Amelinckx}},\ }\href {\doibase https://doi.org/10.1016/0025-5408(72)90133-X}
  {\bibfield  {journal} {\bibinfo  {journal} {Mat. Res. Bull.}\ }\textbf
  {\bibinfo {volume} {7}},\ \bibinfo {pages} {831 } (\bibinfo {year}
  {1972})}\BibitemShut {NoStop}%
\bibitem [{\citenamefont {Polking}\ \emph {et~al.}(2011)\citenamefont
  {Polking}, \citenamefont {Zheng}, \citenamefont {Ramesh},\ and\ \citenamefont
  {Alivisatos}}]{PolkingGeTeDW}%
  \BibitemOpen
  \bibfield  {author} {\bibinfo {author} {\bibfnamefont {M.~J.}\ \bibnamefont
  {Polking}}, \bibinfo {author} {\bibfnamefont {H.}~\bibnamefont {Zheng}},
  \bibinfo {author} {\bibfnamefont {R.}~\bibnamefont {Ramesh}}, \ and\ \bibinfo
  {author} {\bibfnamefont {A.~P.}\ \bibnamefont {Alivisatos}},\ }\href
  {\doibase 10.1021/ja108309s} {\bibfield  {journal} {\bibinfo  {journal} {J.
  Am. Chem. Soc.}\ }\textbf {\bibinfo {volume} {133}},\ \bibinfo {pages} {2044}
  (\bibinfo {year} {2011})}\BibitemShut {NoStop}%
\bibitem [{\citenamefont {Kim}\ and\ \citenamefont {Lee}(2019)}]{Kim2019}%
  \BibitemOpen
  \bibfield  {author} {\bibinfo {author} {\bibfnamefont {S.}~\bibnamefont
  {Kim}}\ and\ \bibinfo {author} {\bibfnamefont {H.~S.}\ \bibnamefont {Lee}},\
  }\href {\doibase 10.1007/s12540-018-0194-4} {\bibfield  {journal} {\bibinfo
  {journal} {Met. Mater. Int.}\ }\textbf {\bibinfo {volume} {25}},\ \bibinfo
  {pages} {528} (\bibinfo {year} {2019})}\BibitemShut {NoStop}%
\bibitem [{\citenamefont {Kriegner}\ \emph {et~al.}(2019)\citenamefont
  {Kriegner}, \citenamefont {Springholz}, \citenamefont {Richter},
  \citenamefont {Pilet}, \citenamefont {M{\"u}ller}, \citenamefont {Capron},
  \citenamefont {Berger}, \citenamefont {Hol{\'y}}, \citenamefont {Dil},\ and\
  \citenamefont {Krempask{\'y}}}]{KrignerGeTeDW}%
  \BibitemOpen
  \bibfield  {author} {\bibinfo {author} {\bibfnamefont {D.}~\bibnamefont
  {Kriegner}}, \bibinfo {author} {\bibfnamefont {G.}~\bibnamefont
  {Springholz}}, \bibinfo {author} {\bibfnamefont {C.}~\bibnamefont {Richter}},
  \bibinfo {author} {\bibfnamefont {N.}~\bibnamefont {Pilet}}, \bibinfo
  {author} {\bibfnamefont {E.}~\bibnamefont {M{\"u}ller}}, \bibinfo {author}
  {\bibfnamefont {M.}~\bibnamefont {Capron}}, \bibinfo {author} {\bibfnamefont
  {H.}~\bibnamefont {Berger}}, \bibinfo {author} {\bibfnamefont
  {V.}~\bibnamefont {Hol{\'y}}}, \bibinfo {author} {\bibfnamefont {J.~H.}\
  \bibnamefont {Dil}}, \ and\ \bibinfo {author} {\bibfnamefont
  {J.}~\bibnamefont {Krempask{\'y}}},\ }\href {\doibase 10.3390/cryst9070335}
  {\bibfield  {journal} {\bibinfo  {journal} {Crystals}\ }\textbf {\bibinfo
  {volume} {9}},\ \bibinfo {pages} {335} (\bibinfo {year} {2019})}\BibitemShut
  {NoStop}%
\bibitem [{\citenamefont {Catalan}\ \emph {et~al.}(2012)\citenamefont
  {Catalan}, \citenamefont {Seidel}, \citenamefont {Ramesh},\ and\
  \citenamefont {Scott}}]{RevModPhysNanoDom}%
  \BibitemOpen
  \bibfield  {author} {\bibinfo {author} {\bibfnamefont {G.}~\bibnamefont
  {Catalan}}, \bibinfo {author} {\bibfnamefont {J.}~\bibnamefont {Seidel}},
  \bibinfo {author} {\bibfnamefont {R.}~\bibnamefont {Ramesh}}, \ and\ \bibinfo
  {author} {\bibfnamefont {J.~F.}\ \bibnamefont {Scott}},\ }\href {\doibase
  10.1103/RevModPhys.84.119} {\bibfield  {journal} {\bibinfo  {journal} {Rev.
  Mod. Phys.}\ }\textbf {\bibinfo {volume} {84}},\ \bibinfo {pages} {119}
  (\bibinfo {year} {2012})}\BibitemShut {NoStop}%
\bibitem [{\citenamefont {Seidel}\ \emph {et~al.}(2009)\citenamefont {Seidel},
  \citenamefont {Martin}, \citenamefont {He}, \citenamefont {Zhan},
  \citenamefont {Chu}, \citenamefont {Rother}, \citenamefont {Hawkridge},
  \citenamefont {Maksymovych}, \citenamefont {Yu}, \citenamefont {Gajek},
  \citenamefont {Balke}, \citenamefont {Kalinin}, \citenamefont {Gemming},
  \citenamefont {Wang}, \citenamefont {Catalan}, \citenamefont {Scott},
  \citenamefont {Spaldin}, \citenamefont {Orenstein},\ and\ \citenamefont
  {Ramesh}}]{MainBiFeO3}%
  \BibitemOpen
  \bibfield  {author} {\bibinfo {author} {\bibfnamefont {J.}~\bibnamefont
  {Seidel}}, \bibinfo {author} {\bibfnamefont {L.~W.}\ \bibnamefont {Martin}},
  \bibinfo {author} {\bibfnamefont {Q.}~\bibnamefont {He}}, \bibinfo {author}
  {\bibfnamefont {Q.}~\bibnamefont {Zhan}}, \bibinfo {author} {\bibfnamefont
  {Y.-H.}\ \bibnamefont {Chu}}, \bibinfo {author} {\bibfnamefont
  {A.}~\bibnamefont {Rother}}, \bibinfo {author} {\bibfnamefont {M.~E.}\
  \bibnamefont {Hawkridge}}, \bibinfo {author} {\bibfnamefont {P.}~\bibnamefont
  {Maksymovych}}, \bibinfo {author} {\bibfnamefont {P.}~\bibnamefont {Yu}},
  \bibinfo {author} {\bibfnamefont {M.}~\bibnamefont {Gajek}}, \bibinfo
  {author} {\bibfnamefont {N.}~\bibnamefont {Balke}}, \bibinfo {author}
  {\bibfnamefont {S.~V.}\ \bibnamefont {Kalinin}}, \bibinfo {author}
  {\bibfnamefont {S.}~\bibnamefont {Gemming}}, \bibinfo {author} {\bibfnamefont
  {F.}~\bibnamefont {Wang}}, \bibinfo {author} {\bibfnamefont {G.}~\bibnamefont
  {Catalan}}, \bibinfo {author} {\bibfnamefont {J.~F.}\ \bibnamefont {Scott}},
  \bibinfo {author} {\bibfnamefont {N.~A.}\ \bibnamefont {Spaldin}}, \bibinfo
  {author} {\bibfnamefont {J.}~\bibnamefont {Orenstein}}, \ and\ \bibinfo
  {author} {\bibfnamefont {R.}~\bibnamefont {Ramesh}},\ }\href
  {http://dx.doi.org/10.1038/nmat2373} {\bibfield  {journal} {\bibinfo
  {journal} {Nat. Mater.}\ }\textbf {\bibinfo {volume} {8}},\ \bibinfo {pages}
  {229} (\bibinfo {year} {2009})}\BibitemShut {NoStop}%
\bibitem [{\citenamefont {Choi}\ \emph {et~al.}(2010)\citenamefont {Choi},
  \citenamefont {Horibe}, \citenamefont {Yi}, \citenamefont {Choi},
  \citenamefont {Wu},\ and\ \citenamefont {Cheong}}]{InsulatingYMnO3}%
  \BibitemOpen
  \bibfield  {author} {\bibinfo {author} {\bibfnamefont {T.}~\bibnamefont
  {Choi}}, \bibinfo {author} {\bibfnamefont {Y.}~\bibnamefont {Horibe}},
  \bibinfo {author} {\bibfnamefont {H.~T.}\ \bibnamefont {Yi}}, \bibinfo
  {author} {\bibfnamefont {Y.~J.}\ \bibnamefont {Choi}}, \bibinfo {author}
  {\bibfnamefont {W.}~\bibnamefont {Wu}}, \ and\ \bibinfo {author}
  {\bibfnamefont {S.-W.}\ \bibnamefont {Cheong}},\ }\href
  {http://dx.doi.org/10.1038/nmat2632} {\bibfield  {journal} {\bibinfo
  {journal} {Nat. Mater.}\ }\textbf {\bibinfo {volume} {9}},\ \bibinfo {pages}
  {253} (\bibinfo {year} {2010})}\BibitemShut {NoStop}%
\bibitem [{\citenamefont {Farokhipoor}\ and\ \citenamefont
  {Noheda}(2011)}]{BiFeO3PRL}%
  \BibitemOpen
  \bibfield  {author} {\bibinfo {author} {\bibfnamefont {S.}~\bibnamefont
  {Farokhipoor}}\ and\ \bibinfo {author} {\bibfnamefont {B.}~\bibnamefont
  {Noheda}},\ }\href {\doibase 10.1103/PhysRevLett.107.127601} {\bibfield
  {journal} {\bibinfo  {journal} {Phys. Rev. Lett.}\ }\textbf {\bibinfo
  {volume} {107}},\ \bibinfo {pages} {127601} (\bibinfo {year}
  {2011})}\BibitemShut {NoStop}%
\bibitem [{\citenamefont {Jin}\ \emph {et~al.}(2018)\citenamefont {Jin},
  \citenamefont {Xiao}, \citenamefont {Yang}, \citenamefont {Zhang},
  \citenamefont {Lu}, \citenamefont {Chu}, \citenamefont {Cheong},
  \citenamefont {Li}, \citenamefont {Kan}, \citenamefont {Yue}, \citenamefont
  {Li}, \citenamefont {Ju}, \citenamefont {Huang},\ and\ \citenamefont
  {Zhu}}]{CondTTAPL}%
  \BibitemOpen
  \bibfield  {author} {\bibinfo {author} {\bibfnamefont {Y.}~\bibnamefont
  {Jin}}, \bibinfo {author} {\bibfnamefont {S.}~\bibnamefont {Xiao}}, \bibinfo
  {author} {\bibfnamefont {J.-C.}\ \bibnamefont {Yang}}, \bibinfo {author}
  {\bibfnamefont {J.}~\bibnamefont {Zhang}}, \bibinfo {author} {\bibfnamefont
  {X.}~\bibnamefont {Lu}}, \bibinfo {author} {\bibfnamefont {Y.-H.}\
  \bibnamefont {Chu}}, \bibinfo {author} {\bibfnamefont {S.-W.}\ \bibnamefont
  {Cheong}}, \bibinfo {author} {\bibfnamefont {J.}~\bibnamefont {Li}}, \bibinfo
  {author} {\bibfnamefont {Y.}~\bibnamefont {Kan}}, \bibinfo {author}
  {\bibfnamefont {C.}~\bibnamefont {Yue}}, \bibinfo {author} {\bibfnamefont
  {Y.}~\bibnamefont {Li}}, \bibinfo {author} {\bibfnamefont {C.}~\bibnamefont
  {Ju}}, \bibinfo {author} {\bibfnamefont {F.}~\bibnamefont {Huang}}, \ and\
  \bibinfo {author} {\bibfnamefont {J.}~\bibnamefont {Zhu}},\ }\href {\doibase
  10.1063/1.5045721} {\bibfield  {journal} {\bibinfo  {journal} {Appl. Phys.
  Lett.}\ }\textbf {\bibinfo {volume} {113}},\ \bibinfo {pages} {082904}
  (\bibinfo {year} {2018})}\BibitemShut {NoStop}%
\bibitem [{\citenamefont {Chen}\ \emph {et~al.}(2018)\citenamefont {Chen},
  \citenamefont {Paillard}, \citenamefont {Zhao}, \citenamefont
  {{\'I}{\~n}iguez}, \citenamefont {Yang},\ and\ \citenamefont
  {Bellaiche}}]{Fermilvlperovskites}%
  \BibitemOpen
  \bibfield  {author} {\bibinfo {author} {\bibfnamefont {L.}~\bibnamefont
  {Chen}}, \bibinfo {author} {\bibfnamefont {C.}~\bibnamefont {Paillard}},
  \bibinfo {author} {\bibfnamefont {H.~J.}\ \bibnamefont {Zhao}}, \bibinfo
  {author} {\bibfnamefont {J.}~\bibnamefont {{\'I}{\~n}iguez}}, \bibinfo
  {author} {\bibfnamefont {Y.}~\bibnamefont {Yang}}, \ and\ \bibinfo {author}
  {\bibfnamefont {L.}~\bibnamefont {Bellaiche}},\ }\href {\doibase
  10.1038/s41524-018-0134-3} {\bibfield  {journal} {\bibinfo  {journal} {npj
  Comput. Mater.}\ }\textbf {\bibinfo {volume} {4}},\ \bibinfo {pages} {75}
  (\bibinfo {year} {2018})}\BibitemShut {NoStop}%
\bibitem [{\citenamefont {Mante}\ and\ \citenamefont
  {Volger}(1971)}]{Mante1971}%
  \BibitemOpen
  \bibfield  {author} {\bibinfo {author} {\bibfnamefont {A.~J.~H.}\
  \bibnamefont {Mante}}\ and\ \bibinfo {author} {\bibfnamefont
  {J.}~\bibnamefont {Volger}},\ }\href {\doibase 10.1016/0031-8914(71)90164-9}
  {\bibfield  {journal} {\bibinfo  {journal} {Physica}\ }\textbf {\bibinfo
  {volume} {52}},\ \bibinfo {pages} {577} (\bibinfo {year} {1971})}\BibitemShut
  {NoStop}%
\bibitem [{\citenamefont {Weilert}\ \emph {et~al.}(1993)\citenamefont
  {Weilert}, \citenamefont {Msall}, \citenamefont {Anderson},\ and\
  \citenamefont {Wolfe}}]{Weilert1993}%
  \BibitemOpen
  \bibfield  {author} {\bibinfo {author} {\bibfnamefont {M.~A.}\ \bibnamefont
  {Weilert}}, \bibinfo {author} {\bibfnamefont {M.~E.}\ \bibnamefont {Msall}},
  \bibinfo {author} {\bibfnamefont {A.~C.}\ \bibnamefont {Anderson}}, \ and\
  \bibinfo {author} {\bibfnamefont {J.~P.}\ \bibnamefont {Wolfe}},\ }\href
  {\doibase 10.1103/PhysRevLett.71.735} {\bibfield  {journal} {\bibinfo
  {journal} {Phys. Rev. Lett.}\ }\textbf {\bibinfo {volume} {71}},\ \bibinfo
  {pages} {735} (\bibinfo {year} {1993})}\BibitemShut {NoStop}%
\bibitem [{\citenamefont {Mielcarek}\ \emph {et~al.}(2001)\citenamefont
  {Mielcarek}, \citenamefont {Mr{\'o}z}, \citenamefont {Tylczy{\'n}ski},
  \citenamefont {Piskunowicz}, \citenamefont {Trybu\l{}a},\ and\ \citenamefont
  {Bromberek}}]{Mielcarek2001}%
  \BibitemOpen
  \bibfield  {author} {\bibinfo {author} {\bibfnamefont {S.}~\bibnamefont
  {Mielcarek}}, \bibinfo {author} {\bibfnamefont {B.}~\bibnamefont {Mr{\'o}z}},
  \bibinfo {author} {\bibfnamefont {Z.}~\bibnamefont {Tylczy{\'n}ski}},
  \bibinfo {author} {\bibfnamefont {P.}~\bibnamefont {Piskunowicz}}, \bibinfo
  {author} {\bibfnamefont {Z.}~\bibnamefont {Trybu\l{}a}}, \ and\ \bibinfo
  {author} {\bibfnamefont {M.}~\bibnamefont {Bromberek}},\ }\href {\doibase
  https://doi.org/10.1016/S0921-4526(00)00584-6} {\bibfield  {journal}
  {\bibinfo  {journal} {Physica B}\ }\textbf {\bibinfo {volume} {299}},\
  \bibinfo {pages} {83 } (\bibinfo {year} {2001})}\BibitemShut {NoStop}%
\bibitem [{\citenamefont {Hopkins}\ \emph {et~al.}(2013)\citenamefont
  {Hopkins}, \citenamefont {Adamo}, \citenamefont {Ye}, \citenamefont {Huey},
  \citenamefont {Lee}, \citenamefont {Schlom},\ and\ \citenamefont
  {Ihlefeld}}]{Hopkins2013}%
  \BibitemOpen
  \bibfield  {author} {\bibinfo {author} {\bibfnamefont {P.~E.}\ \bibnamefont
  {Hopkins}}, \bibinfo {author} {\bibfnamefont {C.}~\bibnamefont {Adamo}},
  \bibinfo {author} {\bibfnamefont {L.}~\bibnamefont {Ye}}, \bibinfo {author}
  {\bibfnamefont {B.~D.}\ \bibnamefont {Huey}}, \bibinfo {author}
  {\bibfnamefont {S.~R.}\ \bibnamefont {Lee}}, \bibinfo {author} {\bibfnamefont
  {D.~G.}\ \bibnamefont {Schlom}}, \ and\ \bibinfo {author} {\bibfnamefont
  {J.~F.}\ \bibnamefont {Ihlefeld}},\ }\href {\doibase 10.1063/1.4798497}
  {\bibfield  {journal} {\bibinfo  {journal} {Appl. Phys. Lett.}\ }\textbf
  {\bibinfo {volume} {102}},\ \bibinfo {pages} {121903} (\bibinfo {year}
  {2013})}\BibitemShut {NoStop}%
\bibitem [{\citenamefont {Li}\ \emph {et~al.}(2014)\citenamefont {Li},
  \citenamefont {Ding}, \citenamefont {Ren}, \citenamefont {Moya},
  \citenamefont {Li}, \citenamefont {Sun},\ and\ \citenamefont
  {Salje}}]{Li2014}%
  \BibitemOpen
  \bibfield  {author} {\bibinfo {author} {\bibfnamefont {S.}~\bibnamefont
  {Li}}, \bibinfo {author} {\bibfnamefont {X.}~\bibnamefont {Ding}}, \bibinfo
  {author} {\bibfnamefont {J.}~\bibnamefont {Ren}}, \bibinfo {author}
  {\bibfnamefont {X.}~\bibnamefont {Moya}}, \bibinfo {author} {\bibfnamefont
  {J.}~\bibnamefont {Li}}, \bibinfo {author} {\bibfnamefont {J.}~\bibnamefont
  {Sun}}, \ and\ \bibinfo {author} {\bibfnamefont {E.~K.~H.}\ \bibnamefont
  {Salje}},\ }\href {\doibase 10.1038/srep06375} {\bibfield  {journal}
  {\bibinfo  {journal} {Sci. Rep.}\ }\textbf {\bibinfo {volume} {4}},\ \bibinfo
  {pages} {6375} (\bibinfo {year} {2014})}\BibitemShut {NoStop}%
\bibitem [{\citenamefont {Ihlefeld}\ \emph {et~al.}(2015)\citenamefont
  {Ihlefeld}, \citenamefont {Foley}, \citenamefont {Scrymgeour}, \citenamefont
  {Michael}, \citenamefont {McKenzie}, \citenamefont {Medlin}, \citenamefont
  {Wallace}, \citenamefont {Trolier-McKinstry},\ and\ \citenamefont
  {Hopkins}}]{Ihlefeld2015}%
  \BibitemOpen
  \bibfield  {author} {\bibinfo {author} {\bibfnamefont {J.~F.}\ \bibnamefont
  {Ihlefeld}}, \bibinfo {author} {\bibfnamefont {B.~M.}\ \bibnamefont {Foley}},
  \bibinfo {author} {\bibfnamefont {D.~A.}\ \bibnamefont {Scrymgeour}},
  \bibinfo {author} {\bibfnamefont {J.~R.}\ \bibnamefont {Michael}}, \bibinfo
  {author} {\bibfnamefont {B.~B.}\ \bibnamefont {McKenzie}}, \bibinfo {author}
  {\bibfnamefont {D.~L.}\ \bibnamefont {Medlin}}, \bibinfo {author}
  {\bibfnamefont {M.}~\bibnamefont {Wallace}}, \bibinfo {author} {\bibfnamefont
  {S.}~\bibnamefont {Trolier-McKinstry}}, \ and\ \bibinfo {author}
  {\bibfnamefont {P.~E.}\ \bibnamefont {Hopkins}},\ }\href {\doibase
  10.1021/nl504505t} {\bibfield  {journal} {\bibinfo  {journal} {Nano Lett.}\
  }\textbf {\bibinfo {volume} {15}},\ \bibinfo {pages} {1791} (\bibinfo {year}
  {2015})}\BibitemShut {NoStop}%
\bibitem [{\citenamefont {Seijas-Bellido}\ \emph {et~al.}(2017)\citenamefont
  {Seijas-Bellido}, \citenamefont {Escorihuela-Sayalero}, \citenamefont {Royo},
  \citenamefont {Ljungberg}, \citenamefont {Wojde\l{}}, \citenamefont
  {\'I\~niguez},\ and\ \citenamefont {Rurali}}]{DWThermCond1}%
  \BibitemOpen
  \bibfield  {author} {\bibinfo {author} {\bibfnamefont {J.~A.}\ \bibnamefont
  {Seijas-Bellido}}, \bibinfo {author} {\bibfnamefont {C.}~\bibnamefont
  {Escorihuela-Sayalero}}, \bibinfo {author} {\bibfnamefont {M.}~\bibnamefont
  {Royo}}, \bibinfo {author} {\bibfnamefont {M.~P.}\ \bibnamefont {Ljungberg}},
  \bibinfo {author} {\bibfnamefont {J.~C.}\ \bibnamefont {Wojde\l{}}}, \bibinfo
  {author} {\bibfnamefont {J.}~\bibnamefont {\'I\~niguez}}, \ and\ \bibinfo
  {author} {\bibfnamefont {R.}~\bibnamefont {Rurali}},\ }\href {\doibase
  10.1103/PhysRevB.96.140101} {\bibfield  {journal} {\bibinfo  {journal} {Phys.
  Rev. B}\ }\textbf {\bibinfo {volume} {96}},\ \bibinfo {pages} {140101}
  (\bibinfo {year} {2017})}\BibitemShut {NoStop}%
\bibitem [{\citenamefont {Royo}\ \emph {et~al.}(2017)\citenamefont {Royo},
  \citenamefont {Escorihuela-Sayalero}, \citenamefont {\'I\~niguez},\ and\
  \citenamefont {Rurali}}]{DWThermCond2}%
  \BibitemOpen
  \bibfield  {author} {\bibinfo {author} {\bibfnamefont {M.}~\bibnamefont
  {Royo}}, \bibinfo {author} {\bibfnamefont {C.}~\bibnamefont
  {Escorihuela-Sayalero}}, \bibinfo {author} {\bibfnamefont {J.}~\bibnamefont
  {\'I\~niguez}}, \ and\ \bibinfo {author} {\bibfnamefont {R.}~\bibnamefont
  {Rurali}},\ }\href {\doibase 10.1103/PhysRevMaterials.1.051402} {\bibfield
  {journal} {\bibinfo  {journal} {Phys. Rev. Mater.}\ }\textbf {\bibinfo
  {volume} {1}},\ \bibinfo {pages} {051402} (\bibinfo {year}
  {2017})}\BibitemShut {NoStop}%
\bibitem [{\citenamefont {Foley}\ \emph {et~al.}(2018)\citenamefont {Foley},
  \citenamefont {Wallace}, \citenamefont {Gaskins}, \citenamefont {Paisley},
  \citenamefont {Johnson-Wilke}, \citenamefont {Kim}, \citenamefont {Ryan},
  \citenamefont {Trolier-McKinstry}, \citenamefont {Hopkins},\ and\
  \citenamefont {Ihlefeld}}]{Foley2018}%
  \BibitemOpen
  \bibfield  {author} {\bibinfo {author} {\bibfnamefont {B.~M.}\ \bibnamefont
  {Foley}}, \bibinfo {author} {\bibfnamefont {M.}~\bibnamefont {Wallace}},
  \bibinfo {author} {\bibfnamefont {J.~T.}\ \bibnamefont {Gaskins}}, \bibinfo
  {author} {\bibfnamefont {E.~A.}\ \bibnamefont {Paisley}}, \bibinfo {author}
  {\bibfnamefont {R.~L.}\ \bibnamefont {Johnson-Wilke}}, \bibinfo {author}
  {\bibfnamefont {J.-W.}\ \bibnamefont {Kim}}, \bibinfo {author} {\bibfnamefont
  {P.~J.}\ \bibnamefont {Ryan}}, \bibinfo {author} {\bibfnamefont
  {S.}~\bibnamefont {Trolier-McKinstry}}, \bibinfo {author} {\bibfnamefont
  {P.~E.}\ \bibnamefont {Hopkins}}, \ and\ \bibinfo {author} {\bibfnamefont
  {J.~F.}\ \bibnamefont {Ihlefeld}},\ }\href {\doibase 10.1021/acsami.8b04169}
  {\bibfield  {journal} {\bibinfo  {journal} {ACS Appl. Mater. Interfaces}\
  }\textbf {\bibinfo {volume} {10}},\ \bibinfo {pages} {25493} (\bibinfo {year}
  {2018})}\BibitemShut {NoStop}%
\bibitem [{\citenamefont {Taherinejad}\ \emph {et~al.}(2012)\citenamefont
  {Taherinejad}, \citenamefont {Vanderbilt}, \citenamefont {Marton},
  \citenamefont {Stepkova},\ and\ \citenamefont {Hlinka}}]{Vanderbilt1}%
  \BibitemOpen
  \bibfield  {author} {\bibinfo {author} {\bibfnamefont {M.}~\bibnamefont
  {Taherinejad}}, \bibinfo {author} {\bibfnamefont {D.}~\bibnamefont
  {Vanderbilt}}, \bibinfo {author} {\bibfnamefont {P.}~\bibnamefont {Marton}},
  \bibinfo {author} {\bibfnamefont {V.}~\bibnamefont {Stepkova}}, \ and\
  \bibinfo {author} {\bibfnamefont {J.}~\bibnamefont {Hlinka}},\ }\href
  {\doibase 10.1103/PhysRevB.86.155138} {\bibfield  {journal} {\bibinfo
  {journal} {Phys. Rev. B}\ }\textbf {\bibinfo {volume} {86}},\ \bibinfo
  {pages} {155138} (\bibinfo {year} {2012})}\BibitemShut {NoStop}%
\bibitem [{\citenamefont {Sist}\ \emph {et~al.}(2018)\citenamefont {Sist},
  \citenamefont {Kasai}, \citenamefont {Hedegaard},\ and\ \citenamefont
  {Iversen}}]{GeTeBo}%
  \BibitemOpen
  \bibfield  {author} {\bibinfo {author} {\bibfnamefont {M.}~\bibnamefont
  {Sist}}, \bibinfo {author} {\bibfnamefont {H.}~\bibnamefont {Kasai}},
  \bibinfo {author} {\bibfnamefont {E.~M.~J.}\ \bibnamefont {Hedegaard}}, \
  and\ \bibinfo {author} {\bibfnamefont {B.~B.}\ \bibnamefont {Iversen}},\
  }\href {\doibase 10.1103/PhysRevB.97.094116} {\bibfield  {journal} {\bibinfo
  {journal} {Phys. Rev. B}\ }\textbf {\bibinfo {volume} {97}},\ \bibinfo
  {pages} {094116} (\bibinfo {year} {2018})}\BibitemShut {NoStop}%
\bibitem [{\citenamefont {Lubk}\ \emph {et~al.}(2009)\citenamefont {Lubk},
  \citenamefont {Gemming},\ and\ \citenamefont {Spaldin}}]{DFTBiFeO31}%
  \BibitemOpen
  \bibfield  {author} {\bibinfo {author} {\bibfnamefont {A.}~\bibnamefont
  {Lubk}}, \bibinfo {author} {\bibfnamefont {S.}~\bibnamefont {Gemming}}, \
  and\ \bibinfo {author} {\bibfnamefont {N.~A.}\ \bibnamefont {Spaldin}},\
  }\href {\doibase 10.1103/PhysRevB.80.104110} {\bibfield  {journal} {\bibinfo
  {journal} {Phys. Rev. B}\ }\textbf {\bibinfo {volume} {80}},\ \bibinfo
  {pages} {104110} (\bibinfo {year} {2009})}\BibitemShut {NoStop}%
\bibitem [{\citenamefont {Di\'eguez}\ \emph {et~al.}(2013)\citenamefont
  {Di\'eguez}, \citenamefont {Aguado-Puente}, \citenamefont {Junquera},\ and\
  \citenamefont {\'I\~niguez}}]{Dieguez2013}%
  \BibitemOpen
  \bibfield  {author} {\bibinfo {author} {\bibfnamefont {O.}~\bibnamefont
  {Di\'eguez}}, \bibinfo {author} {\bibfnamefont {P.}~\bibnamefont
  {Aguado-Puente}}, \bibinfo {author} {\bibfnamefont {J.}~\bibnamefont
  {Junquera}}, \ and\ \bibinfo {author} {\bibfnamefont {J.}~\bibnamefont
  {\'I\~niguez}},\ }\href {\doibase 10.1103/PhysRevB.87.024102} {\bibfield
  {journal} {\bibinfo  {journal} {Phys. Rev. B}\ }\textbf {\bibinfo {volume}
  {87}},\ \bibinfo {pages} {024102} (\bibinfo {year} {2013})}\BibitemShut
  {NoStop}%
\bibitem [{\citenamefont {Chen}\ \emph {et~al.}(2017)\citenamefont {Chen},
  \citenamefont {Kuo},\ and\ \citenamefont {Chew}}]{DFTBiFeO32}%
  \BibitemOpen
  \bibfield  {author} {\bibinfo {author} {\bibfnamefont {Y.-W.}\ \bibnamefont
  {Chen}}, \bibinfo {author} {\bibfnamefont {J.-L.}\ \bibnamefont {Kuo}}, \
  and\ \bibinfo {author} {\bibfnamefont {K.-H.}\ \bibnamefont {Chew}},\ }\href
  {\doibase 10.1063/1.4998456} {\bibfield  {journal} {\bibinfo  {journal} {J.
  Appl. Phys.}\ }\textbf {\bibinfo {volume} {122}},\ \bibinfo {pages} {075103}
  (\bibinfo {year} {2017})}\BibitemShut {NoStop}%
\bibitem [{\citenamefont {Gong}\ \emph {et~al.}(2018)\citenamefont {Gong},
  \citenamefont {Li}, \citenamefont {Zhang}, \citenamefont {Li}, \citenamefont
  {Zheng}, \citenamefont {Yang}, \citenamefont {Huang}, \citenamefont {Lin},
  \citenamefont {Yan},\ and\ \citenamefont {Liu}}]{GONG20189}%
  \BibitemOpen
  \bibfield  {author} {\bibinfo {author} {\bibfnamefont {J.}~\bibnamefont
  {Gong}}, \bibinfo {author} {\bibfnamefont {C.}~\bibnamefont {Li}}, \bibinfo
  {author} {\bibfnamefont {Y.}~\bibnamefont {Zhang}}, \bibinfo {author}
  {\bibfnamefont {Y.}~\bibnamefont {Li}}, \bibinfo {author} {\bibfnamefont
  {S.}~\bibnamefont {Zheng}}, \bibinfo {author} {\bibfnamefont
  {K.}~\bibnamefont {Yang}}, \bibinfo {author} {\bibfnamefont {R.}~\bibnamefont
  {Huang}}, \bibinfo {author} {\bibfnamefont {L.}~\bibnamefont {Lin}}, \bibinfo
  {author} {\bibfnamefont {Z.}~\bibnamefont {Yan}}, \ and\ \bibinfo {author}
  {\bibfnamefont {J.-M.}\ \bibnamefont {Liu}},\ }\href {\doibase
  https://doi.org/10.1016/j.mtphys.2018.06.002} {\bibfield  {journal} {\bibinfo
   {journal} {Mater. Today Phys.}\ }\textbf {\bibinfo {volume} {6}},\ \bibinfo
  {pages} {9 } (\bibinfo {year} {2018})}\BibitemShut {NoStop}%
\bibitem [{\citenamefont {Perdew}\ \emph {et~al.}(1996)\citenamefont {Perdew},
  \citenamefont {Burke},\ and\ \citenamefont {Ernzerhof}}]{GGAPBE}%
  \BibitemOpen
  \bibfield  {author} {\bibinfo {author} {\bibfnamefont {J.~P.}\ \bibnamefont
  {Perdew}}, \bibinfo {author} {\bibfnamefont {K.}~\bibnamefont {Burke}}, \
  and\ \bibinfo {author} {\bibfnamefont {M.}~\bibnamefont {Ernzerhof}},\ }\href
  {\doibase 10.1103/PhysRevLett.77.3865} {\bibfield  {journal} {\bibinfo
  {journal} {Phys. Rev. Lett.}\ }\textbf {\bibinfo {volume} {77}},\ \bibinfo
  {pages} {3865} (\bibinfo {year} {1996})}\BibitemShut {NoStop}%
\bibitem [{\citenamefont {Goedecker}\ \emph {et~al.}(1996)\citenamefont
  {Goedecker}, \citenamefont {Teter},\ and\ \citenamefont
  {Hutter}}]{HGHpseudo}%
  \BibitemOpen
  \bibfield  {author} {\bibinfo {author} {\bibfnamefont {S.}~\bibnamefont
  {Goedecker}}, \bibinfo {author} {\bibfnamefont {M.}~\bibnamefont {Teter}}, \
  and\ \bibinfo {author} {\bibfnamefont {J.}~\bibnamefont {Hutter}},\ }\href
  {\doibase 10.1103/PhysRevB.54.1703} {\bibfield  {journal} {\bibinfo
  {journal} {Phys. Rev. B}\ }\textbf {\bibinfo {volume} {54}},\ \bibinfo
  {pages} {1703} (\bibinfo {year} {1996})}\BibitemShut {NoStop}%
\bibitem [{\citenamefont {Gonze}\ \emph {et~al.}(2009)\citenamefont {Gonze},
  \citenamefont {Amadon}, \citenamefont {Anglade}, \citenamefont {Beuken},
  \citenamefont {Bottin}, \citenamefont {Boulanger}, \citenamefont {Bruneval},
  \citenamefont {Caliste}, \citenamefont {Caracas}, \citenamefont
  {C\^{o}t\'{e}}, \citenamefont {Deutsch}, \citenamefont {Genovese},
  \citenamefont {Ghosez}, \citenamefont {Giantomassi}, \citenamefont
  {Goedecker}, \citenamefont {Hamann}, \citenamefont {Hermet}, \citenamefont
  {Jollet}, \citenamefont {Jomard}, \citenamefont {Leroux}, \citenamefont
  {Mancini}, \citenamefont {Mazevet}, \citenamefont {Oliveira}, \citenamefont
  {Onida}, \citenamefont {Pouillon}, \citenamefont {Rangel}, \citenamefont
  {Rignanese}, \citenamefont {Sangalli}, \citenamefont {Shaltaf}, \citenamefont
  {Torrent}, \citenamefont {Verstraete}, \citenamefont {Zerah},\ and\
  \citenamefont {Zwanziger}}]{ABINIT}%
  \BibitemOpen
  \bibfield  {author} {\bibinfo {author} {\bibfnamefont {X.}~\bibnamefont
  {Gonze}}, \bibinfo {author} {\bibfnamefont {B.}~\bibnamefont {Amadon}},
  \bibinfo {author} {\bibfnamefont {P.-M.}\ \bibnamefont {Anglade}}, \bibinfo
  {author} {\bibfnamefont {J.-M.}\ \bibnamefont {Beuken}}, \bibinfo {author}
  {\bibfnamefont {F.}~\bibnamefont {Bottin}}, \bibinfo {author} {\bibfnamefont
  {P.}~\bibnamefont {Boulanger}}, \bibinfo {author} {\bibfnamefont
  {F.}~\bibnamefont {Bruneval}}, \bibinfo {author} {\bibfnamefont
  {D.}~\bibnamefont {Caliste}}, \bibinfo {author} {\bibfnamefont
  {R.}~\bibnamefont {Caracas}}, \bibinfo {author} {\bibfnamefont
  {M.}~\bibnamefont {C\^{o}t\'{e}}}, \bibinfo {author} {\bibfnamefont
  {T.}~\bibnamefont {Deutsch}}, \bibinfo {author} {\bibfnamefont
  {L.}~\bibnamefont {Genovese}}, \bibinfo {author} {\bibfnamefont
  {P.}~\bibnamefont {Ghosez}}, \bibinfo {author} {\bibfnamefont
  {M.}~\bibnamefont {Giantomassi}}, \bibinfo {author} {\bibfnamefont
  {S.}~\bibnamefont {Goedecker}}, \bibinfo {author} {\bibfnamefont
  {D.}~\bibnamefont {Hamann}}, \bibinfo {author} {\bibfnamefont
  {P.}~\bibnamefont {Hermet}}, \bibinfo {author} {\bibfnamefont
  {F.}~\bibnamefont {Jollet}}, \bibinfo {author} {\bibfnamefont
  {G.}~\bibnamefont {Jomard}}, \bibinfo {author} {\bibfnamefont
  {S.}~\bibnamefont {Leroux}}, \bibinfo {author} {\bibfnamefont
  {M.}~\bibnamefont {Mancini}}, \bibinfo {author} {\bibfnamefont
  {S.}~\bibnamefont {Mazevet}}, \bibinfo {author} {\bibfnamefont
  {M.}~\bibnamefont {Oliveira}}, \bibinfo {author} {\bibfnamefont
  {G.}~\bibnamefont {Onida}}, \bibinfo {author} {\bibfnamefont
  {Y.}~\bibnamefont {Pouillon}}, \bibinfo {author} {\bibfnamefont
  {T.}~\bibnamefont {Rangel}}, \bibinfo {author} {\bibfnamefont {G.-M.}\
  \bibnamefont {Rignanese}}, \bibinfo {author} {\bibfnamefont {D.}~\bibnamefont
  {Sangalli}}, \bibinfo {author} {\bibfnamefont {R.}~\bibnamefont {Shaltaf}},
  \bibinfo {author} {\bibfnamefont {M.}~\bibnamefont {Torrent}}, \bibinfo
  {author} {\bibfnamefont {M.}~\bibnamefont {Verstraete}}, \bibinfo {author}
  {\bibfnamefont {G.}~\bibnamefont {Zerah}}, \ and\ \bibinfo {author}
  {\bibfnamefont {J.}~\bibnamefont {Zwanziger}},\ }\href {\doibase
  https://doi.org/10.1016/j.cpc.2009.07.007} {\bibfield  {journal} {\bibinfo
  {journal} {Comput. Phys. Commun.}\ }\textbf {\bibinfo {volume} {180}},\
  \bibinfo {pages} {2582 } (\bibinfo {year} {2009})}\BibitemShut {NoStop}%
\bibitem [{\citenamefont {Gonze}\ \emph {et~al.}(2016)\citenamefont {Gonze},
  \citenamefont {Jollet}, \citenamefont {Araujo}, \citenamefont {Adams},
  \citenamefont {Amadon}, \citenamefont {Applencourt}, \citenamefont {Audouze},
  \citenamefont {Beuken}, \citenamefont {Bieder}, \citenamefont {Bokhanchuk},
  \citenamefont {Bousquet}, \citenamefont {Bruneval}, \citenamefont {Caliste},
  \citenamefont {C\^{o}t{\'e}}, \citenamefont {Dahm}, \citenamefont {Pieve},
  \citenamefont {Delaveau}, \citenamefont {Gennaro}, \citenamefont {Dorado},
  \citenamefont {Espejo}, \citenamefont {Geneste}, \citenamefont {Genovese},
  \citenamefont {Gerossier}, \citenamefont {Giantomassi}, \citenamefont
  {Gillet}, \citenamefont {Hamann}, \citenamefont {He}, \citenamefont {Jomard},
  \citenamefont {Janssen}, \citenamefont {Roux}, \citenamefont {Levitt},
  \citenamefont {Lherbier}, \citenamefont {Liu}, \citenamefont
  {Luka\v{c}evi{\'c}}, \citenamefont {Martin}, \citenamefont {Martins},
  \citenamefont {Oliveira}, \citenamefont {Ponc{\'e}}, \citenamefont
  {Pouillon}, \citenamefont {Rangel}, \citenamefont {Rignanese}, \citenamefont
  {Romero}, \citenamefont {Rousseau}, \citenamefont {Rubel}, \citenamefont
  {Shukri}, \citenamefont {Stankovski}, \citenamefont {Torrent}, \citenamefont
  {Setten}, \citenamefont {Troeye}, \citenamefont {Verstraete}, \citenamefont
  {Waroquiers}, \citenamefont {Wiktor}, \citenamefont {Xu}, \citenamefont
  {Zhou},\ and\ \citenamefont {Zwanziger}}]{ABINIT2}%
  \BibitemOpen
  \bibfield  {author} {\bibinfo {author} {\bibfnamefont {X.}~\bibnamefont
  {Gonze}}, \bibinfo {author} {\bibfnamefont {F.}~\bibnamefont {Jollet}},
  \bibinfo {author} {\bibfnamefont {F.~A.}\ \bibnamefont {Araujo}}, \bibinfo
  {author} {\bibfnamefont {D.}~\bibnamefont {Adams}}, \bibinfo {author}
  {\bibfnamefont {B.}~\bibnamefont {Amadon}}, \bibinfo {author} {\bibfnamefont
  {T.}~\bibnamefont {Applencourt}}, \bibinfo {author} {\bibfnamefont
  {C.}~\bibnamefont {Audouze}}, \bibinfo {author} {\bibfnamefont {J.-M.}\
  \bibnamefont {Beuken}}, \bibinfo {author} {\bibfnamefont {J.}~\bibnamefont
  {Bieder}}, \bibinfo {author} {\bibfnamefont {A.}~\bibnamefont {Bokhanchuk}},
  \bibinfo {author} {\bibfnamefont {E.}~\bibnamefont {Bousquet}}, \bibinfo
  {author} {\bibfnamefont {F.}~\bibnamefont {Bruneval}}, \bibinfo {author}
  {\bibfnamefont {D.}~\bibnamefont {Caliste}}, \bibinfo {author} {\bibfnamefont
  {M.}~\bibnamefont {C\^{o}t{\'e}}}, \bibinfo {author} {\bibfnamefont
  {F.}~\bibnamefont {Dahm}}, \bibinfo {author} {\bibfnamefont {F.~D.}\
  \bibnamefont {Pieve}}, \bibinfo {author} {\bibfnamefont {M.}~\bibnamefont
  {Delaveau}}, \bibinfo {author} {\bibfnamefont {M.~D.}\ \bibnamefont
  {Gennaro}}, \bibinfo {author} {\bibfnamefont {B.}~\bibnamefont {Dorado}},
  \bibinfo {author} {\bibfnamefont {C.}~\bibnamefont {Espejo}}, \bibinfo
  {author} {\bibfnamefont {G.}~\bibnamefont {Geneste}}, \bibinfo {author}
  {\bibfnamefont {L.}~\bibnamefont {Genovese}}, \bibinfo {author}
  {\bibfnamefont {A.}~\bibnamefont {Gerossier}}, \bibinfo {author}
  {\bibfnamefont {M.}~\bibnamefont {Giantomassi}}, \bibinfo {author}
  {\bibfnamefont {Y.}~\bibnamefont {Gillet}}, \bibinfo {author} {\bibfnamefont
  {D.}~\bibnamefont {Hamann}}, \bibinfo {author} {\bibfnamefont
  {L.}~\bibnamefont {He}}, \bibinfo {author} {\bibfnamefont {G.}~\bibnamefont
  {Jomard}}, \bibinfo {author} {\bibfnamefont {J.~L.}\ \bibnamefont {Janssen}},
  \bibinfo {author} {\bibfnamefont {S.~L.}\ \bibnamefont {Roux}}, \bibinfo
  {author} {\bibfnamefont {A.}~\bibnamefont {Levitt}}, \bibinfo {author}
  {\bibfnamefont {A.}~\bibnamefont {Lherbier}}, \bibinfo {author}
  {\bibfnamefont {F.}~\bibnamefont {Liu}}, \bibinfo {author} {\bibfnamefont
  {I.}~\bibnamefont {Luka\v{c}evi{\'c}}}, \bibinfo {author} {\bibfnamefont
  {A.}~\bibnamefont {Martin}}, \bibinfo {author} {\bibfnamefont
  {C.}~\bibnamefont {Martins}}, \bibinfo {author} {\bibfnamefont
  {M.}~\bibnamefont {Oliveira}}, \bibinfo {author} {\bibfnamefont
  {S.}~\bibnamefont {Ponc{\'e}}}, \bibinfo {author} {\bibfnamefont
  {Y.}~\bibnamefont {Pouillon}}, \bibinfo {author} {\bibfnamefont
  {T.}~\bibnamefont {Rangel}}, \bibinfo {author} {\bibfnamefont {G.-M.}\
  \bibnamefont {Rignanese}}, \bibinfo {author} {\bibfnamefont {A.}~\bibnamefont
  {Romero}}, \bibinfo {author} {\bibfnamefont {B.}~\bibnamefont {Rousseau}},
  \bibinfo {author} {\bibfnamefont {O.}~\bibnamefont {Rubel}}, \bibinfo
  {author} {\bibfnamefont {A.}~\bibnamefont {Shukri}}, \bibinfo {author}
  {\bibfnamefont {M.}~\bibnamefont {Stankovski}}, \bibinfo {author}
  {\bibfnamefont {M.}~\bibnamefont {Torrent}}, \bibinfo {author} {\bibfnamefont
  {M.~V.}\ \bibnamefont {Setten}}, \bibinfo {author} {\bibfnamefont {B.~V.}\
  \bibnamefont {Troeye}}, \bibinfo {author} {\bibfnamefont {M.}~\bibnamefont
  {Verstraete}}, \bibinfo {author} {\bibfnamefont {D.}~\bibnamefont
  {Waroquiers}}, \bibinfo {author} {\bibfnamefont {J.}~\bibnamefont {Wiktor}},
  \bibinfo {author} {\bibfnamefont {B.}~\bibnamefont {Xu}}, \bibinfo {author}
  {\bibfnamefont {A.}~\bibnamefont {Zhou}}, \ and\ \bibinfo {author}
  {\bibfnamefont {J.}~\bibnamefont {Zwanziger}},\ }\href {\doibase
  https://doi.org/10.1016/j.cpc.2016.04.003} {\bibfield  {journal} {\bibinfo
  {journal} {Comput. Phys. Commun.}\ }\textbf {\bibinfo {volume} {205}},\
  \bibinfo {pages} {106 } (\bibinfo {year} {2016})}\BibitemShut {NoStop}%
\bibitem [{\citenamefont {Marzari}(1996)}]{MarzariPhD}%
  \BibitemOpen
  \bibfield  {author} {\bibinfo {author} {\bibfnamefont {N.}~\bibnamefont
  {Marzari}},\ }\emph {\bibinfo {title} {Ab-initio molecular dynamics for
  metallic systems}},\ \href
  {http://inis.iaea.org/search/search.aspx?orig_q=RN:30045813; British Library
  Document Supply Centre- DSC:D203390} {Ph.D. thesis} (\bibinfo {year}
  {1996})\BibitemShut {NoStop}%
\bibitem [{Note1()}]{Note1}%
  \BibitemOpen
  \bibinfo {note} {For 39$^{\circ }$ tail-to-tail domain wall, polarization
  values along the trigonal axis in one of the domains were taken as negative
  to obtain the $\protect \qopname \relax o{tanh}(x)$ dependence of
  polarization, see Fig.~\ref {fig2}(a).}\BibitemShut {Stop}%
\bibitem [{\citenamefont {Meyer}\ and\ \citenamefont
  {Vanderbilt}(2002)}]{Vanderbiltpbti03}%
  \BibitemOpen
  \bibfield  {author} {\bibinfo {author} {\bibfnamefont {B.}~\bibnamefont
  {Meyer}}\ and\ \bibinfo {author} {\bibfnamefont {D.}~\bibnamefont
  {Vanderbilt}},\ }\href {\doibase 10.1103/PhysRevB.65.104111} {\bibfield
  {journal} {\bibinfo  {journal} {Phys. Rev. B}\ }\textbf {\bibinfo {volume}
  {65}},\ \bibinfo {pages} {104111} (\bibinfo {year} {2002})}\BibitemShut
  {NoStop}%
\bibitem [{\citenamefont {Dangi\ifmmode~\acute{c}\else \'{c}\fi{}}\ \emph
  {et~al.}(2018)\citenamefont {Dangi\ifmmode~\acute{c}\else \'{c}\fi{}},
  \citenamefont {Murphy}, \citenamefont {Murray}, \citenamefont {Fahy},\ and\
  \citenamefont {Savi\ifmmode~\acute{c}\else \'{c}\fi{}}}]{Mi}%
  \BibitemOpen
  \bibfield  {author} {\bibinfo {author} {\bibfnamefont {{\DJ}.}~\bibnamefont
  {Dangi\ifmmode~\acute{c}\else \'{c}\fi{}}}, \bibinfo {author} {\bibfnamefont
  {A.~R.}\ \bibnamefont {Murphy}}, \bibinfo {author} {\bibfnamefont
  {{\'E}.~D.}\ \bibnamefont {Murray}}, \bibinfo {author} {\bibfnamefont
  {S.}~\bibnamefont {Fahy}}, \ and\ \bibinfo {author} {\bibfnamefont
  {I.}~\bibnamefont {Savi\ifmmode~\acute{c}\else \'{c}\fi{}}},\ }\href
  {\doibase 10.1103/PhysRevB.97.224106} {\bibfield  {journal} {\bibinfo
  {journal} {Phys. Rev. B}\ }\textbf {\bibinfo {volume} {97}},\ \bibinfo
  {pages} {224106} (\bibinfo {year} {2018})}\BibitemShut {NoStop}%
\bibitem [{\citenamefont {Chattopadhyay}\ \emph {et~al.}(1987)\citenamefont
  {Chattopadhyay}, \citenamefont {Boucherle},\ and\ \citenamefont
  {vonSchnering}}]{main}%
  \BibitemOpen
  \bibfield  {author} {\bibinfo {author} {\bibfnamefont {T.}~\bibnamefont
  {Chattopadhyay}}, \bibinfo {author} {\bibfnamefont {J.~X.}\ \bibnamefont
  {Boucherle}}, \ and\ \bibinfo {author} {\bibfnamefont {H.~G.}\ \bibnamefont
  {vonSchnering}},\ }\href {http://stacks.iop.org/0022-3719/20/i=10/a=012}
  {\bibfield  {journal} {\bibinfo  {journal} {J. Phys. C: Solid State Phys.}\
  }\textbf {\bibinfo {volume} {20}},\ \bibinfo {pages} {1431} (\bibinfo {year}
  {1987})}\BibitemShut {NoStop}%
\bibitem [{\citenamefont {Chatterji}\ \emph {et~al.}(2015)\citenamefont
  {Chatterji}, \citenamefont {Kumar},\ and\ \citenamefont {Wdowik}}]{newmain}%
  \BibitemOpen
  \bibfield  {author} {\bibinfo {author} {\bibfnamefont {T.}~\bibnamefont
  {Chatterji}}, \bibinfo {author} {\bibfnamefont {C.~M.~N.}\ \bibnamefont
  {Kumar}}, \ and\ \bibinfo {author} {\bibfnamefont {U.~D.}\ \bibnamefont
  {Wdowik}},\ }\href {\doibase 10.1103/PhysRevB.91.054110} {\bibfield
  {journal} {\bibinfo  {journal} {Phys. Rev. B}\ }\textbf {\bibinfo {volume}
  {91}},\ \bibinfo {pages} {054110} (\bibinfo {year} {2015})}\BibitemShut
  {NoStop}%
\bibitem [{\citenamefont {Hanus}\ \emph {et~al.}(2018)\citenamefont {Hanus},
  \citenamefont {Garg},\ and\ \citenamefont {Snyder}}]{Hanus2018}%
  \BibitemOpen
  \bibfield  {author} {\bibinfo {author} {\bibfnamefont {R.}~\bibnamefont
  {Hanus}}, \bibinfo {author} {\bibfnamefont {A.}~\bibnamefont {Garg}}, \ and\
  \bibinfo {author} {\bibfnamefont {G.~J.}\ \bibnamefont {Snyder}},\ }\href
  {\doibase 10.1038/s42005-018-0070-z} {\bibfield  {journal} {\bibinfo
  {journal} {Commun. Phys.}\ }\textbf {\bibinfo {volume} {1}},\ \bibinfo
  {pages} {78} (\bibinfo {year} {2018})}\BibitemShut {NoStop}%
\bibitem [{\citenamefont {Swartz}\ and\ \citenamefont {Pohl}(1989)}]{AMM}%
  \BibitemOpen
  \bibfield  {author} {\bibinfo {author} {\bibfnamefont {E.~T.}\ \bibnamefont
  {Swartz}}\ and\ \bibinfo {author} {\bibfnamefont {R.~O.}\ \bibnamefont
  {Pohl}},\ }\href {\doibase 10.1103/RevModPhys.61.605} {\bibfield  {journal}
  {\bibinfo  {journal} {Rev. Mod. Phys.}\ }\textbf {\bibinfo {volume} {61}},\
  \bibinfo {pages} {605} (\bibinfo {year} {1989})}\BibitemShut {NoStop}%
\bibitem [{\citenamefont {Seijas-Bellido}\ \emph {et~al.}(2018)\citenamefont
  {Seijas-Bellido}, \citenamefont {Aramberri}, \citenamefont {\'I\~niguez},\
  and\ \citenamefont {Rurali}}]{DWThermCond}%
  \BibitemOpen
  \bibfield  {author} {\bibinfo {author} {\bibfnamefont {J.~A.}\ \bibnamefont
  {Seijas-Bellido}}, \bibinfo {author} {\bibfnamefont {H.}~\bibnamefont
  {Aramberri}}, \bibinfo {author} {\bibfnamefont {J.}~\bibnamefont
  {\'I\~niguez}}, \ and\ \bibinfo {author} {\bibfnamefont {R.}~\bibnamefont
  {Rurali}},\ }\href {\doibase 10.1103/PhysRevB.97.184306} {\bibfield
  {journal} {\bibinfo  {journal} {Phys. Rev. B}\ }\textbf {\bibinfo {volume}
  {97}},\ \bibinfo {pages} {184306} (\bibinfo {year} {2018})}\BibitemShut
  {NoStop}%
\end{thebibliography}%

\end{document}